\documentclass[a4paper]{article}

\usepackage[svgnames,x11names,table]{xcolor}
\definecolor{xlinkcolor}{cmyk}{1,1,0,0}
\usepackage[colorlinks=true,urlcolor=xlinkcolor,citecolor=xlinkcolor,linkcolor=xlinkcolor,bookmarks=true]{hyperref}
\definecolor{orcidlogocol}{HTML}{A6CE39}
\usepackage[sort&compress, round, comma, authoryear]{natbib}
\setcitestyle{citesep={;}}
\usepackage{arxiv}
\usepackage[utf8]{inputenc} % allow utf-8 input
\usepackage[T1]{fontenc}    % use 8-bit T1 fonts
\usepackage{url}            % simple URL typesetting
\usepackage{booktabs}       % professional-quality tables
\usepackage{amsfonts}       % blackboard math symbols
\usepackage{nicefrac}       % compact symbols for 1/2, etc.
\usepackage{microtype}      % microtypography
\usepackage{graphicx}
\usepackage{natbib}
\usepackage{doi}
\usepackage{authblk}
\usepackage{amsmath}
\usepackage{subcaption}
\usepackage{caption}
\usepackage{geometry}
\usepackage{multicol}
\usepackage{cleveref}
\usepackage{booktabs}
\usepackage{array}
\usepackage{moresize}
\usepackage{txfonts}
\usepackage{sectsty}
\usepackage{etoolbox}
\usepackage{orcidlink}
\usepackage{enumitem}
\setlist[itemize]{leftmargin=*}
\usepackage[font={small,sl},labelfont=bf]{caption}
\usepackage{float}
\usepackage{adjustbox}
\usepackage{abstract}

\subsectionfont{\normalfont\itshape}

\makeatletter

\renewcommand\maketitle
  {\noindent
   {\Large\bfseries\@title}%
   \medskip\par\noindent
   {\large\bfseries\@author}%
   \hfill
   {\large\@date}%
   \bigskip\par\noindent
  }
\makeatother

\title {\centering \textsf{An algorithm to calculate the relative orbit, ephemeris, and individual masses of unresolved astrometric binaries} \\[1.25ex] \fontsize{11}{12.5}\selectfont \textsf{\centering Example of application on the newest Gaia DR3 binaries: the ESMORGA catalog}\\[1ex]}

\author[1,2,3]{\textcolor{xlinkcolor} {\hspace{1.9cm} Xabier Pérez-Couto\thanks{xabier.perez.couto@usc.es}\,\orcidlink{0000-0001-5797-252X} }}
\author[1,2,4]{\textcolor{xlinkcolor}{Jose Á. Docobo\thanks{joseangel.docobo@usc.es}\,\orcidlink{0000-0001-9556-8624}}}
\author[2]{\textcolor{xlinkcolor}{Pedro P. Campo\thanks{pedropablo.campo@usc.es}\,\orcidlink{0000-0001-7268-5879}}}

\affil[1]{\footnotesize{CITMAga, 15782 Santiago de Compostela, Galiza, Spain}}
\affil[2]{Observatorio Astronómico R. M. Aller (OARMA), Universidade de Santiago de Compostela (USC), Campus Vida, 15782 Santiago de Compostela, Galiza, Spain}
\affil[3]{\footnotesize{Agrupación Astronómica Coruñesa ``Ío'', 15005 A Coruña, Galiza, Spain}}
\affil[4]{Real Academia de Ciencias de Zaragoza, Facultad de Ciencias, C/ Pedro Cerbuna 12, E-50009 Zaragoza, Spain}

\date{}

\renewcommand{\headeright}{A draft version June 28, 2023}
\renewcommand{\shorttitle}{\scriptsize{\textit{arXiv} Template}}

\begin{document}

\twocolumn[
  \begin{@twocolumnfalse}

    \maketitle

    \begin{abstract}
	The recent Gaia Data Release 3 has unveiled a catalog of over eight hundred thousand binary systems, providing orbital solutions for half of them. Since most of them are unresolved astrometric binaries, several astrophysical parameters that can be only derived from their relative orbits together with spectroscopic data, such as the individual stellar masses, remain unknown. Indeed, only the mass of the primary, $\texttt{m1}$, and a wide interval, $\texttt{[m2\_lower, m2\_upper]}$, for the secondary companion of main-sequence astrometric binaries have been derived to date \citep{hiddentreasure}. In order to obtain the correct values for each component, we propose an analytic algorithm to estimate the two most probable relative orbits and magnitude differences of a certain main-sequence or subgiant astrometric binary using all available Gaia data. Subsequently, both possible solutions are constrained to the one that is consistent with $\texttt{m1, m2\_lower}$ and $\texttt{m2\_upper}$. Moreover, we deduce not only the correct values of the individual masses for each binary but also the size of the telescope necessary to resolve their components. The workflow of our algorithm as well as the ESMORGA (Ephemeris, Stellar Masses, and relative ORbits from GAia) catalog with more than one hundred thousand individual masses, spectral types, and effective temperatures derivated from its application are also presented.
\end{abstract}
  \end{@twocolumnfalse}\vspace{0.1cm}
]
\saythanks

% keywords can be removed
\keywords{binaries: astrometric \and stars: fundamental parameters \and methods: data analysis \and methods: numerical}

\section{Introduction}\label{introduction}

Since the orbital motion of binary stars was discovered around 200 years ago \citep{herschel}, the study of stellar systems has become one of the most outstanding topics in astronomy. Their importance lies mainly, in the calculation of binary orbits because they allow the stellar masses to be directly determined, through the well-known Kepler's Third Law, 
\begin{equation}\label{thirdkeplerlaw}
    \mathcal{M}_A + \mathcal{M}_B = \dfrac{a^3}{P^2},
\end{equation} and the quotient of masses $q=\frac{\mathcal{M}_B}{\mathcal{M}_A}$, being the most accurate method to date. However, beyond the stellar dynamics, the scope of binary research extends to other recent areas of astronomy such as stellar evolution, exoplanetology and galaxy mapping, since the stellar mass is the parameter that best tells us how stars will evolve and die, and provides us with a good distance estimation tool: the dynamical parallax:
\begin{equation}\label{dynamicalparallax}
    \pi'' = \dfrac{a''}{a},
\end{equation}
where $a$ is the semi-major axis of the relative orbit measured in astronomical units and $a''$ is that but calculated in arc seconds. Moreover, if there are some planets orbiting around them, their physical properties will be expressed as functions of the stellar parameters, mainly of the mass. 

For this reason, instrumentation aiming to allow the detection of this systems, together with a large variety of mathematical methods to compute their orbits, were developed in the last century, such as the analytic methods of Thiele-Innes-Van den Bos \citep{heintz}, Cid \citep{cid, cid1960}, Docobo \citep{docobo1985, docobo2012}, and different graphical methods \citep{zwiers, vidal}, among others. Among the observational techniques that can be found, we have to highlight speckle interferometry \citep{mcallister}, adaptive optics and space surveys \citep{docoboandrade2015} as the cutting-edge techniques that allow more orbits to be calculated. In this work, we focus on the latest data release of the Gaia mission, the Gaia DR3. 

The high accuracy in positions and proper motions of Gaia has allowed to detect an unprecedented quantity of astrometric binary stars (338 215 systems, up to now), multiplying by six the number of binary systems observed from ground-based measurements and computing the orbits of half of them.
The orbital solutions provided by Gaia are expressed as seven orbital elements, three of Campbell ($P,T,e$) and the four Innes constants ($A, B, F, G$). However, the four remaining Campbell orbital parameters can be recovered just by solving the system of equations below \citep{halbwachs}:

\begin{flalign}
A &= \alpha''(\cos \omega \cos \Omega - \sin \omega \sin \Omega \cos I) \label{A} \\
B &= \alpha''(\cos \omega \sin \Omega + \sin \omega \cos \Omega \cos I) \label{B} \\
F &= \alpha''(- \sin \omega \cos \Omega - \cos \omega \sin \Omega \cos I) \label{F} \\
G &= \alpha''(- \sin \omega \sin \Omega + \cos \omega \cos \Omega \cos I) \label{G} 
\end{flalign}

%C_1 &= a_1 \sin{I} \sin{\omega_1}\\
%H_1 &= a_1 \sin{I} \cos{\omega_1}

Through the Gaia $\texttt{gaiadr3.binary\_masses}$ an estimation for individual masses, \texttt{m1} and \texttt{m2}, of the components of an astrometric binary are provided if it also has a spectroscopic orbit (as it occurs for 17578 \texttt{AstroSpectroSB1} solutions), since the astrometric orbit only shows the movement of the photocenter of the system around its center of mass instead of the orbital motion of a component around the other, from which $a''$ is derived. Otherwise, for pure astrometric orbits (111 792 of the total), only a wide interval, $\texttt{[m2\_lower, m2\_upper]}$, for the secondary mass, is bounded, by assuming an isochrone-derived primary mass, $\texttt{m1}$ \citep{hiddentreasure}. Hence, to obtain both individual masses, $\texttt{m1}$, and $\texttt{m2}$, of the new Gaia DR3 astrometric binaries, we have developed an algorithm for transforming the photocentric orbits calculated by Gaia into relative orbits. From them, individual masses are derived, saving consistency with the constraints imposed by $\texttt{gaiadr3.binary\_masses}$ solutions to achieve concrete and robust solutions for both stellar masses. Finally, our program calculates the ephemeris of the apparent orbit, and therefore, we can determine the angular separation at given dates to assess the possibility of resolving them. The algorithm also provides us with the magnitude difference between the components, and consequently the minimum aperture needed to resolve a given system comes to light.

This manuscript is organized as follows: in section \ref{methodoloy} we present our methodology, from the basic definitions and the theoretical framework to the in-depth description of the algorithm. The analysis of the results obtained by applying our methodology to an actual validation dataset as well as a study on the numerical convergence of the algorithm is shown in section \ref{results} . Finally, some conclusions are commented in the Discussion section and, after the Bibliography, several appendices with the used calibration tables and catalogs obtained through the algorithm are provided.
\section{Methodology}
\label{methodoloy}

\subsection{Theoretical framework}

As commented before, the orbital solutions given by Gaia DR3 NSS astrometric solution table correspond to the movement of the photocenter around the center of mass. This astrometric orbit is related to the relative orbit as follows.
Let $f$ be the mass ratio defined as
\begin{flalign}\label{f}
    f = \dfrac{\mathcal{M}_B}{\mathcal{M}_A + \mathcal{M}_B}.
\end{flalign}
It is well known (see \citet{heintz}, Chapter 22) that if $\textbf{d}$ is the vector distance between the primary and secondary components, then the vector distance from the center of mass ($cm$) to the primary component will be $-f\textbf{d}$ and the one from the $cm$ to the secondary component, $(1-f)\textbf{d}$. In other words, $f$ is the normalized distance from the primary component to the $cm$. On the other hand, the position of the photocenter can be determined through the flux ratio, $\beta$, defined as
\begin{flalign}\label{beta}
    \beta &= \dfrac{f_{V,B}}{f_{V,A} + f_{V,B}} = \dfrac{1}{1+10^{0.4\Delta m}},
\end{flalign}
where $f_{V,A}$ and $f_{V,B}$ are the spectral flux densities of both components. 

\begin{figure}[ht]
    \includegraphics[scale=0.47]{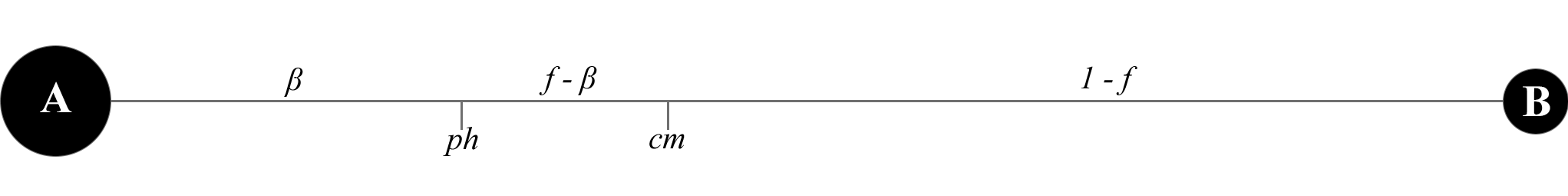}
    \caption{Geometric interpretation of the mass, $f$, and flux, $\beta$, ratios together with the photocenter $ph$ and the center of mass $cm$ of a binary system, with unit distance $d$. Personal contribution based on the original scheme of \citet{heintz}.}
    \label{f-beta} 
\end{figure}

Now, analogously to $f$, $-\beta \textbf{d}$ is the vector that goes from the photocenter ($ph$) to the primary component and $(1-\beta)\textbf{d}$ goes from $ph$ to the secondary star. Consequently, $f-\beta$ is the scale factor of the distance that measures the separation between $cm$ and $ph$ in which the astrometric orbital motion is measured and, therefore, is also the scale factor of the photocentric semi-major axis, $\alpha$, to the relative, $a''$, that is:
\begin{flalign}\label{photocentricsma}
\alpha = a'' \cdot (f - \beta)
\end{flalign}
In Figure \ref{f-beta}, the commented geometric interpretation of $f$ and $\beta$ is visually shown.

Therefore, starting with the $\alpha$ given by the \texttt{gaiadr3.nss\_two\_body\_orbit} table as the semi-major axis of the astrometric orbit (called \texttt{Orbital}), we can compute the semi-major axis of the relative orbit, $a''$, if we know $f$ and $\beta$ or by using \eqref{f} and \eqref{beta}, the individual masses, $\mathcal{M}_A$ and $\mathcal{M}_B$, of both companions, as well as their magnitude difference, $\Delta m$. In the work of \citet{equationssystem} the system of the equations \eqref{thirdkeplerlaw}-\eqref{dynamicalparallax} and \eqref{photocentricsma} given as follows is proposed to be solved for $a''$, and the mass of one of the components if the other mass is known as well as $\Delta m$:
\begin{flalign}\label{eq:equationssystem}
    \begin{cases}
     \mathcal{M}_A + \mathcal{M}_B = \dfrac{a''^3}{\pi''^3 P^2}\\
     \alpha=a''\cdot \left(\dfrac{\mathcal{M}_B}{\mathcal{M}_A + \mathcal{M}_B}-\dfrac{1}{1+10^{0.4\Delta m}}\right),
    \end{cases}
\end{flalign}

However, those stellar parameters are not trivial to calculate, especially if it is a close binary which, as we will see in section \ref{results}, is the most common case of systems detected astrometrically in Gaia DR3. In fact, the calculation of the masses, as commented in section \ref{introduction}, is the main motivation for binary orbital calculation, and therefore one of the last parameters to be obtained in the related processes, by using $a$, $P$, and the quotient of masses, $q$ (or, equivalently, the mass ratio $f$). 

As the reader may notice, two of the last mentioned variables are \textit{a priori} unknown (in truth, they are among the unknowns that we want to solve). Moreover, the magnitude difference is also unknown when the binary cannot be optically resolved, either because it has a large $\Delta m$ or because its components are too close.

Fortunately, for some luminosity classes, there are several calibrations that we can use to obtain the individual luminosities or, equivalently, the absolute magnitudes, and through a Mass-Luminosity Relation (MLR), their individual masses, if we know the spectral type of each component; for example the \citet{pecautmamajek} calibration for main-sequence stars or that of \citet{straizyskuriliene} for subgiants. Very often, even in resolved binaries, the spectral type of each component is not determined, only the combination of spectral features of both components, known as the composite or combined spectrum.

There is a wide variety of methods to separate the composite spectrum,, taking advantage of orbital geometry (e.g. in eclipsing binaries) \citep{griffinetgriffin}, when the radial velocity is high enough so that the individual spectral features are well separated and easy to identify \citep{ferluga,hadrava} or when composite spectrums at different orbital phases are available \citep{simon}. In this work we use the \citet{edwards} through the \citet{campo} implementation as we only need the composite spectrum in its Morgan-Keenan (MK) designation and the magnitude difference of the system to determine the individual spectrum of each companion. Indeed, we need again to know $\Delta m$ to work but, now, we have it as an input parameter in both the Edward process and in the system of equations \eqref{eq:equationssystem}, so that we can take advantage of that to develop an analytical and recursive algorithm (see section \ref{algorithm}) to get the best solution for $\Delta m$ and, thus, the rest of the output parameters. But first, let us introduce how the Edwards process works.

\subsubsection{The Edwards process}\label{edwardsprocess}

The theoretical basis of this method, previously proposed by \citet{christywalker}, is based on thinking about the composite spectrum as a linear interpolation of the individual spectral types, weighted by their luminosities. \citet{edwards} suggested to solve this calculation by means of the following system of equations:
\begin{flalign}\label{eq:edwards}
    \begin{cases}
    S(A) + xS(B) = (1+x)S(A+B)\\
    M[S(A)] - M[S(B)] = \Delta m = -2.5\log{x},
    \end{cases}
\end{flalign}
Here $\{S(i),\:i=A,B\}$ represents the MK spectral types of the component $i$, $S(A+B)$ means the composite MK spectral type and $\{M[S(i)],\:i=A,B\}$ is a luminosity calibration of the $\{S(i),\:i=A,B\}$ MK spectral type.

Neverthless, the problem arises when we have to numerically interpret $S(i)$ with the aim of performing the needed interpolation. Whereas a practical approach that works fine is to replace $S(i)$ by the visual absolute magnitude of the same companion ($M_{Vi}$) so that the first equation in \eqref{eq:edwards} is substituted by
\begin{flalign}\label{edwardsapaño}
    M_{VA} + xM_{VB} = (1+x)M_V,
\end{flalign}
with $M_V$ being the absolute magnitude related to the composite spectrum, this calculus has not any known physical sense.

A more realistic alternative approach is that suggested by \citet{beaverscook} taking the monochromatic normalized flux of the composite spectrum as a weighted-normalized sum of the individual fluxes in the wavelength, $\lambda$, having a similar expression to \eqref{edwardsapaño} but with a more straightforward physical sense:
\begin{flalign}\label{monochromaticnormalizedflux}
F_{\lambda} = \dfrac{F_{\lambda, A} + kF_{\lambda,B}}{1+k},\: \: k\in\mathbb{R}
\end{flalign}
Now, by means of the normalized flux definition for the visual band \citep{campo}
\begin{flalign}
    F_V=\dfrac{f_V}{F},
\end{flalign}
where $f_V$ is the spectral flux density in the visual band and $F$ is the total flux integrated with respect to the wavelength, we can now rewrite \eqref{monochromaticnormalizedflux} as \citep{campo}:
\begin{flalign}\label{eq:newstepflux}
    \dfrac{f_{\lambda}}{F} = \dfrac{\frac{f_{\lambda, A}}{F_A} + k\frac{f_{\lambda,B}}{F_B}}{1+k},\: \: k\in\mathbb{R}
\end{flalign}
For the next step we must use the following definitions of visual and bolometric magnitude as a function of their corresponding fluxes:
\begin{flalign}\label{eq:visualmag}
    m_V=-2.5\log{f_V}\\
    m_{Bol}=-2.5\log{F}
\end{flalign}
In this way, we can invert the logarithms in the previous expressions and rewrite \eqref{eq:newstepflux} as a function of observable data such as visual and bolometric apparent magnitudes \citep{campo}:
\begin{flalign}\label{eq:newstepmag}
    (1+k)\dfrac{10^{-0.4m_V}}{10^{-0.4m_{Bol}}}=\dfrac{10^{-0.4m_{VA}}}{10^{-0.4m_{BolA}}} + k\dfrac{10^{-0.4m_{VB}}}{10^{-0.4m_{BolB}}},\:\: k\in\mathbb{R}
\end{flalign}
However, it is well-known that apparent $m$ and absolute $M$ magnitudes are related to the distance, $d$, in parsecs by means of the distance-modulus expression
\begin{flalign}\label{eq:distancemodulus}
    M - m = 5-5\log{d},
\end{flalign}
so if we replace the visual and bolometric apparent magnitudes in \eqref{eq:newstepmag} by their corresponding forms of \eqref{eq:distancemodulus}, the right side of the previous equation cancels out between the numerator and denominator of each fraction and we have the same relation as a function of the absolute magnitudes \citep{campo}:
\begin{flalign}\label{eq:newstepabsmag}
    (1+k)\dfrac{10^{-0.4M_V}}{10^{-0.4M_{Bol}}}=\dfrac{10^{-0.4M_{VA}}}{10^{-0.4M_{BolA}}} + k\dfrac{10^{-0.4M_{VB}}}{10^{-0.4M_{BolB}}},\:\: k\in\mathbb{R}
\end{flalign}
Finally, it is enough to take the definition of the bolometric correction ($BC$),
\begin{flalign}\label{eq:bc}
    BC = M_{Bol} - M_V,
\end{flalign}
to simplify \eqref{eq:newstepabsmag}:
\begin{flalign}
(1+k)10^{0.4BC} = 10^{0.4BC_A} + k10^{0.4BC_B},\:\:k\in\mathbb{R}
\end{flalign}
Here the $BC$ of the system can be obtained from the composite spectrum by means of some calibration like those of \citet{pecautmamajek} and \citet{straizyskuriliene}, and $k=x=10^{-0.4\Delta m}$ since what \citet{campo} did was to express the Edwards process in terms of the bolometric correction. Furthermore, by considering the $BC$ of each component as a function of their absolute visual magnitudes ($BC_i=\Phi(M_{Vi}),\:i=A,B$), we have the final equation
\begin{flalign}\label{finaledwardsequation}
    (1+k)10^{0.4BC} = 10^{0.4\Phi(M_{VA})} + k10^{0.4\Phi(M_{VA} + \Delta m)}
\end{flalign}
that can be numerically resolved by Campo's variation of the Edwards process using an interpolation of $\phi$ to get $M_{VA}$ and, straight away, $M_{VB}=M_{VA}+\Delta m$ \citep{campo}.

\subsection{The algorithm, step by step}\label{algorithm}

After introducing the theoretical framework in which the methodology of this work is based on, we are ready to present the main original contribution of this work: an analytic algorithm used, given a Gaia astrometric binary, to transform its photocentric semi-major axis, $\alpha$, into the semi-major axis of the relative orbit, $a''$. During the process, the magnitude difference, $\Delta m$, of the unresolved system is derived along with a precise estimate of the individual mass of each component.

Moreover, this algorithm gives us an even more powerful tool: once we have achieved $a''$, we can compute the orbital ephemeris in relative polar coordinates of each system $\{(r_i, v_i),\:i\in \mathbb{N}\}$ and, transforming them into their apparent coordinates $\{(\rho''_i, \theta_i),\:i\in\mathbb{N}\}$, we can derive straightforwardly the minimum $\rho''_{min}$ and maximum $\rho''_{max}$ angular separation between their components. Subsequently, and together with the magnitude difference, we can know for each unresolved astrometric binary the minimum aperture a ground-based telescope will need to resolve it.

\subsubsection{Sample selection and preprocessing}\label{initialsample}

Given a Non-Single Star astrometric or astrospectroscopic solution from Gaia DR3, first we have to determine if it corresponds to a main-sequence star or subgiant, as the calibrations we will use in the next step is only valid for stars whose spectral type is between O3V and L2V or between O5IV and K1IV, respectively. Determining the luminosity class of the input star can be done directly by cross-matching it with the WDS, as it contain the spectral class of a reasonably large sample of binaries. However, we can also estimate the luminosity class of a star, or at least to know if it belongs to those luminosity classes, by means of the logarithm of its surface gravity, $\log{g}$, given in the Gaia DR3 source catalog as \texttt{logg\_gspphot} (in $\log{(cm\cdot s^{-2})}$). For instance, based on the works of \citet{angelov}, \citet{zboril}, \citet{bastien} and  we can constrain the input main-sequence stars to those with $\log{g}>4.0\:dex$ and subgiants if $3.6<=\log{g}<=4.0\:dex$. However, the surface gravity of an unresolved binary thought as a single star is not very reliable, as the cooler companion produce an underestimation of the effective temperature, and thus also of the $\log{g}$ since it is highly covariant with $T_{eff}$ and metallicity \citep{elbadry2018, ting2017}. Hence, instead of using the gravity surface computed by GSP-Phot assuming a single-star model, we have used the Multiple Star Classifier (MSC) individual gravity surfaces, $\texttt{log1\_msc}$ and $\texttt{log2\_msc}$, through a model that assumes a binary system with both components in the same evolutionary stage and with a flux ratio smaller than 5. Subsequently, we redefine with practical purposes the $\log{g}$ as the minima of them, since we want to consider also the subgiant branch. With that aim, the following ADQL query is used to extract the full sample:

\vspace{-0.1cm}

\fontsize{7.7}{10}
\begin{verbatim}
select nss.*, gs.*, bm.*, ap.*
from gaiadr3.nss_two_body_orbit as nss,
gaiadr3.gaia_source as gs,
gaiadr3.binary_masses as bm,
gaiadr3.astrophysical_parameters as ap
where gs.source_id=nss.source_id
and bm.source_id=nss.source_id
and ap.source_id=nss.source_id
and (nss.nss_solution_type=`Orbital'
or nss.nss_solution_type=`Orbital'
or nss.nss_solution_type=`OrbitalAlternative'
or nss.nss_solution_type=`OrbitalAlternativeValidated'
or nss.nss_solution_type=`OrbitalTargetedSearch'
or nss.nss_solution_type=`OrbitalTargetedSearchValidated'
or nss.nss_solution_type=`AstroSpectroSB1')
and ap.logg_msc1>=3.6
and ap.logg_msc2>=3.6
\end{verbatim}

\vspace{-0.1cm}

\normalsize
Next, the seven Campbell orbital elements for the astrometric orbit are derived by using the relations \eqref{A}, \eqref{B}, \eqref{F} and \eqref{G} as well as their uncertainties, as fully detailed in \citet{halbwachs}. Thereafter, we applied the same filter of \citet{hiddentreasure} for discarding low signal-to-noise ratio ($SNR$) solutions, by imposing $SNR > 5$ for $\alpha$ and $SNR > 2$ for $\sin{I}$ in astrometric orbits, while $log_{10}{(SNR)} > 3.7 - 1.1 \log_{10}{P}$ is also required for \texttt{OrbitalAlternative} solutions. Moreover, $SNR > 5$ for $\sqrt{C_1^2+H_1^2}$ and $a_1$ in the astrospectroscopic ones is asked.

Now, for the sake of clarity, let's see how the algorithm works step by step.

\subsubsection{First step: composite spectrum determination} 

Now, for each unresolved binary, we take its composite spectrum as the spectral type of the Gaia source. To do this, we get the absolute magnitude $M_V$ of the system from the $G$-band mean apparent magnitude, $m_G$, given by Gaia DR3 as \texttt{phot\_g\_mean\_mag} ($mag$).To calculate it we only need the parallax (\texttt{parallax}), $G_{BP} - G_{RP}$ color (\texttt{bp-rp}), and the extinction coefficients for both \texttt{bp-rp} (\texttt{ebpminrp\_gspphot}) and \texttt{phot\_g\_mean\_mag} (\texttt{ag\_msc}).

First, the apparent magnitude is corrected from interstellar absorption and scattering by subtracting the extinction in $G$ band available from MSC as \texttt{ag\_msc} ($mag$).
Then, to transform $m_G$ band magnitude into Johnson-Cousins $V$ band, just use the polynomial of degree 3 given by the coefficients shown in table 5.9 of the ``Gaia DR3 documentation - \textit{Relationships with other photometric systems}'' \citep{gaiadocumentation}, 
\begin{flalign}
    G - m_V =& -0.02704 + 0.01424Z - 0.2156Z^2\\ &+ 0.01426Z^3\notag,
\end{flalign}
where $Z = (G_{BP} - G_{RP})_0$ is the extinction-corrected color index. For some sources we found that \texttt{ebpminrp\_gspphot} is not available. For those few cases an extinction of $=0.0807\:mag$ was considered, since that was the median value for the color index extinction over a random sample of 100 sources.

Subsequently, just using \texttt{parallax} ($mas$) and the distance-modulus relation \eqref{eq:distancemodulus} we can obtain the absolute visual magnitude $M_V$ of the system. 

Finally, using the calibrations previously commented for each case, we can assign the composite spectral type as a function of $M_V$ by means of a spline cubic interpolation.
    
\subsubsection{Second step: disentangling the composite spectrum} 

Now, choosing a small enough initial value for the magnitude difference, e.g. $\Delta m_0 = 0.1\: mag$, we can already run the Campo's implementation of the Edwards process, described in subsection \ref{edwardsprocess}. The method uses either the calibration of \citet{pecautmamajek} or that of \citet{straizyskuriliene} to compute the bolometric correction of the input star. Then, by using the initial $\Delta m$, the program solves numerically the non-linear equation \eqref{finaledwardsequation} by means of the Newton-Raphson's method or Nelder-Mead's one (if the first fails). 
    
In this manner, we get the individual absolute visual magnitudes $M_{VA}$, $M_{VB}$ of the companions and, by interpolation, the spectroscopic individual masses, $\mathcal{M}_{SpA}$ and $\mathcal{M}_{SpB}$, are derived.
    
\subsubsection{Third step: initial solution} 

With the given initial $\Delta m_0$, we can calculate the flux ratio $\beta$ by means of the second equality in \eqref{beta}. At the same time, the mass ratio $f$ is computed from \eqref{f} together with the spectroscopic masses, $\mathcal{M}_{SpA}$ and $\mathcal{M}_{SpB}$, estimated in the second step. Therefore, by replacing this two quantities and the photocentric semi-major axis $\alpha$ given by Gaia in expression \eqref{photocentricsma}, we obtain a value for the semi-major axis $a''$ that, like $\alpha$, is measured in arc seconds. This is not a problem because we have the Gaia parallax, $\pi''$, so that by using \eqref{dynamicalparallax} the value of $a$ in astronomical units is recovered.

From the value of $a$ derived in the previous step in addition to the orbital period $P$ available in the Gaia DR3 NSS astrometric solution, we can determine, by means of \eqref{thirdkeplerlaw}, the sum of the masses (called in this case ``orbital masses'', to differentiate them from the spectroscopic ones), $\mathcal{M}_T=\mathcal{M}_A + \mathcal{M}_B$. On the other hand, adding up the spectroscopic masses, we get $\mathcal{M}_{SpT}=\mathcal{M}_{SpA} + \mathcal{M}_{SpB}$. 

\subsubsection{Fourth step: again and again}

It is clear that, if both masses were calculated correctly; that is, if the arbitrary assumed value for $\Delta m$ would be correct, then $\mathcal{M}_T=\mathcal{M}_{SpT}$. For this reason, the core idea of the algorithm is to perform the previous set of calculations with different values of $\Delta m$, until the above equality is verified.

To apply this, what the algorithm does is to repeat the previous steps increasing the magnitude difference by $0.1\:mag$, that is, for the iteration $i$, $\Delta_{mi} = \Delta_{mi-1} + 0.1$. By doing this recursively and performing the following multiplication with the total masses obtained after each iteration,
\begin{flalign}\label{convergence}
    (\mathcal{M}_{SpTi} - \mathcal{M}_{Ti})
(\mathcal{M}_{SpTi-1} - \mathcal{M}_{Ti-1}),
\end{flalign}
it is clear that, by using the Bolzano's Theorem, if the subtractions in \eqref{convergence} are continuous functions, the multiplication \eqref{convergence} will change its sign when the solution, that is, the value of $\Delta m$ for which $\mathcal{M}_T=\mathcal{M}_{SpT}$, is located in the interval $\left[\Delta_{mi-1}, \Delta_{mi}\right]$ (see section \ref{numericalconvergence} for further details).

Then the process continues until an arbitrarily high magnitude difference, e.g. $\Delta m=10\: mag$. Then, the recursive process ends. Subsequently, taking the solution intervals $\left[\Delta_{mi-1}, \Delta_{mi}\right]$ of longitude $0.1\:mag$, we can consider their medium points, $\Delta m = (\Delta_{mi-1}+\Delta_{mi})/2$, as estimators of the final solutions. 

\subsubsection{Fifth step: ephemeris calculation} 

Using the obtained solutions for $\Delta m$ and the mean value of the individual masses (corresponding to the extremes of the interval in which we found the solution) we get, through \eqref{photocentricsma}, the solution for the relative semi-major axis, $a''$.
    
From it, and together with the rest of the orbital parameters derived from the Innes constants, we are in a position to compute the ephemeris of the apparent orbit. Firstly, we calculate the mean anomaly $M$ for an epoch $t$ \citep{cursoastronomia},
\begin{flalign}
    M = \dfrac{2\pi}{P}(t-T)
\end{flalign}
so the eccentric anomaly $E$ can be numerically resolved from the Kepler's equation \citep{cursoastronomia}:
\begin{flalign}\label{keplerequation}
    M = E - e\sin{E}
\end{flalign}
Now, we can calculate the true anomaly, $v$, using the well-know relation \citep{cursoastronomia}
\begin{flalign}\label{trueanomaly}
    \tan{v/2} = \sqrt{(1+e)(1-e)}\tan{E/2}
\end{flalign}
as well as the angular distance $r''$ through the polar form of the ellipse equation \citep{cursoastronomia}:
\begin{flalign}\label{requation}
    r''=\dfrac{a''(1-e^2)}{1+e\cos{v}},
\end{flalign}
being $a''=a\cdot \pi''$.
    
Finally, if we take the following well-known relations between the relative orbit and the apparent one \citep{cursoastronomia, ahmad},
\begin{flalign}\label{relativetoapparent}
    \rho'' \cos{(\theta - \Omega)}=r''\cos{(\omega + v)}\\
    \rho'' \sin{(\theta - \Omega)}=r''\sin{(\omega + v)}\cos{I}
\end{flalign}
and divide the second equation by the first one, we can rewrite them as follows \citep{cursoastronomia, ahmad}:
\begin{flalign}\label{thetarelative}
\tan{(\theta - \Omega)} = \tan{(\omega + v)}\cos{I}
\end{flalign}
\vspace{-0.7cm}
\begin{flalign}\label{rhorelative}
    \rho'' = r''\dfrac{\cos{(\omega + v)}}{\cos{(\theta -\Omega)}}
\end{flalign}
By solving the above system of equations we can obtain the corresponding $\theta$ and $\rho''$. However, it should be noticed that, with the aim of obtaining the minimum $\rho''_{min}$ and maximum $\rho''_{max}$ separations, as wall as in order to be capable of drawing the projected-in-the-sky orbit, it is enough to give values for $\theta_i \in \left[0^{\circ}, 360^{\circ}\right)$ in \eqref{thetarelative}, to calculate $v$ and then, by replacing those last two quantities in \eqref{rhorelative}, we finally get the corresponding $\rho''_i$ for each $\theta_i$ and, thus, $\rho''_{min}$ and $\rho''_{max}$ \citep{cursoastronomia, ahmad}.

\subsection{Finding the true solution}

Up to now, we have been working with a sample of 134598 astrometric orbits for which Gaia only provides the classic four Innes elements. However, for 33467 of them, Gaia also computed a single-lined spectroscopic orbit. In those cases, an extra solution called \texttt{AstroSpectroSB1} is available with two more Innes constants, $C_1$ and $H_1$, from which $a_1 \sin{I}$ and $\omega_1$ can be derived, as explained in \citet{halbwachs}. 

A very interesting quantity that can be obtained from those elements is the binary mass function ($f_m$, in $\mathcal{M}_{\odot}$) that can be calculated from Gaia data as follows \citep{hiddentreasure}:
\begin{equation}
    f_m = (C_1^2 + H_1^2)^{3/2} P^{-2},
\end{equation}
where $C_1$ and $H_1$ are expressed in $au$, and $P$ in years.

Independently, we can compute the binary mass function by means of the individual masses obtained with the algorithm ($\hat{f}_{m}$) for each solution, that is:
\begin{equation}
    \hat{f}_{m} = \dfrac{\mathcal{M}_{SpB}^3\sin^3{I}}{(\mathcal{M}_{SpA} + \mathcal{M}_{SpB})^2}
\end{equation}
Therefore, if the relative semi-major axis computed by our algorithm is correct, then $f_m \approx \hat{f}_{m}$. This provide us with a good test to check which solution if the correct one and, at the same time, to validate it with independent data.

To see if a given solution passes the test, we have propagated the uncertainties through the algorithm and obtained the errors ($\sigma_{f_m}$ and $\sigma_{\hat{f}_{m}}$) of the binary mass functions. 

Hence, we will consider that a certain solution pass the tests if, and only if, the following condition is met:
\begin{equation}
    |f_m - \hat{f}_{m}| \leq \sigma_{f_m} + \sigma_{\hat{f}_{m}}
\end{equation}

Regarding the large data set of \texttt{Orbital} solutions and as we have introduced in Section \ref{introduction}, we can use the \texttt{gaiadr3.binary\_masses} to bound the correct solution among the in-average two solutions provided by our algorithm. In fact, by simply comparing the two pairs of possible individual masses $\{\mathcal{M}_{SpA}^j, \mathcal{M}_{SpB}^j\}\: _{j=1,2}$ for a certain source, we select the correct pair $j$ as the one fulfilling: $\mathcal{M}_{SpA}^j \approx \texttt{m1}$ and $\mathcal{M}_{SpB}^j \in \left[\texttt{m2\_lower},\, \texttt{m2\_upper}\right]$.

\section{Results and discussion} \label{results}

\subsection{On the numerical convergence}\label{numericalconvergence}
It is easy to see that the iterative process of finding solutions described above is equivalent to finding the zeros of the following function:
\begin{equation}\label{deltamathcal}
    \Delta \mathcal{M}\left(\Delta m\right) = \mathcal{M}_T\left(\Delta m\right) - \mathcal{M}_{SpT}\left(\Delta m\right)
\end{equation}

Perhaps the reader is curious about how $\Delta \mathcal{M}(\Delta m)$ behaves for different $\Delta m$ values, especially in the neighbourhood of the possible solutions/zeros of the function. To shed some light on this issue, we have selected a sample of 796 Gaia DR3 astrometric binaries with their composite spectrum available at \textit{SIMBAD} (\url{http://simbad.u-strasbg.fr/simbad}) showing different cases of convergence and computed $\Delta \mathcal{M}$ for $\Delta m \in \left[0.5,8\right]$ $mag$ with steps of $0.1$ $mag$, so we can plot the graphical representation of $\Delta \mathcal{M}$ around the points where $\Delta\mathcal{M}=0$. Four different convergence scenarios were found, being two the mean and median number of solutions.

\begin{figure*}[ht]
    \centering
    \begin{subfigure}[b]{0.45\textwidth}
    \includegraphics[width=\textwidth]{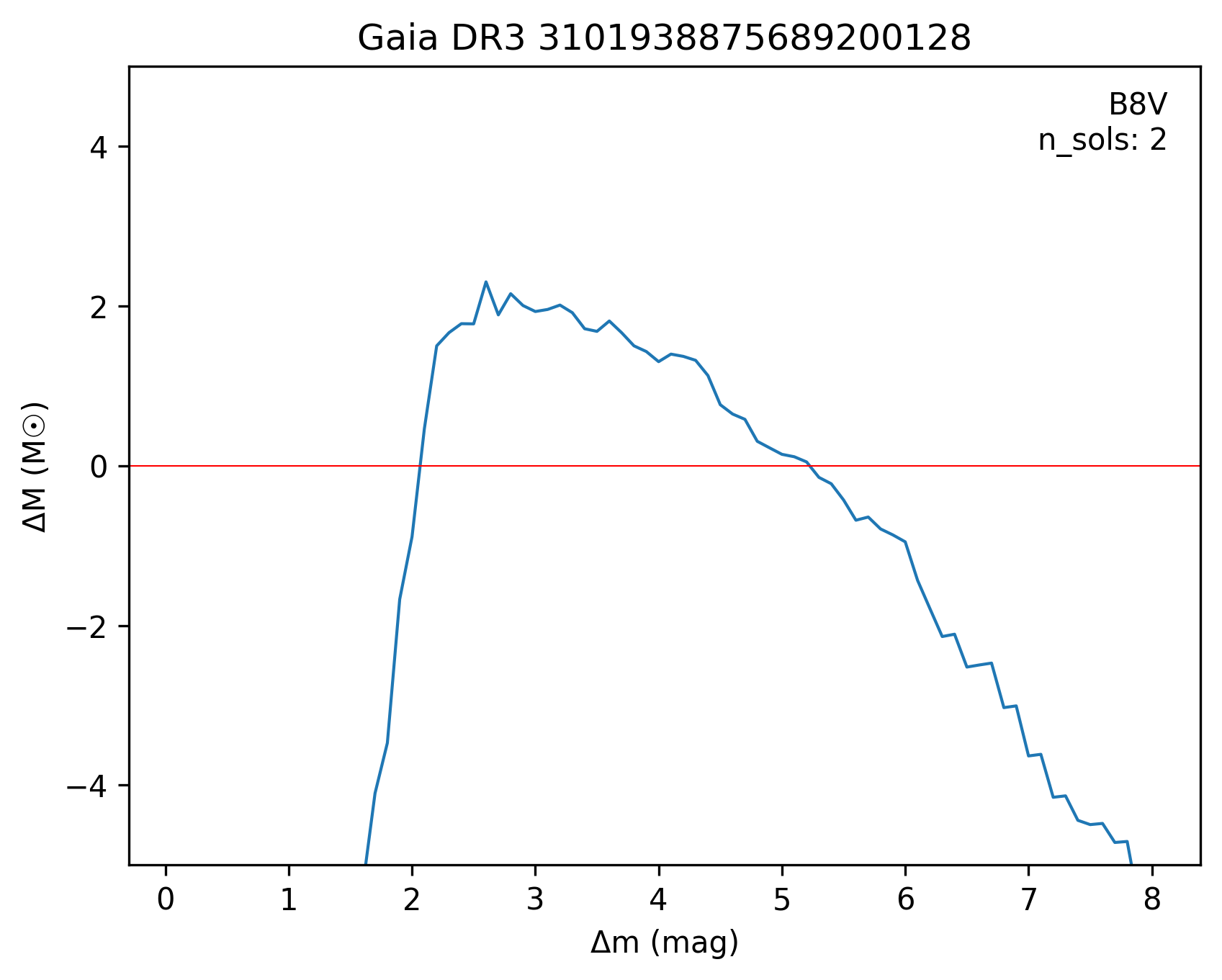}
    \end{subfigure}
     \hfill
    \begin{subfigure}[b]{0.45\textwidth}
    \includegraphics[width=\textwidth]{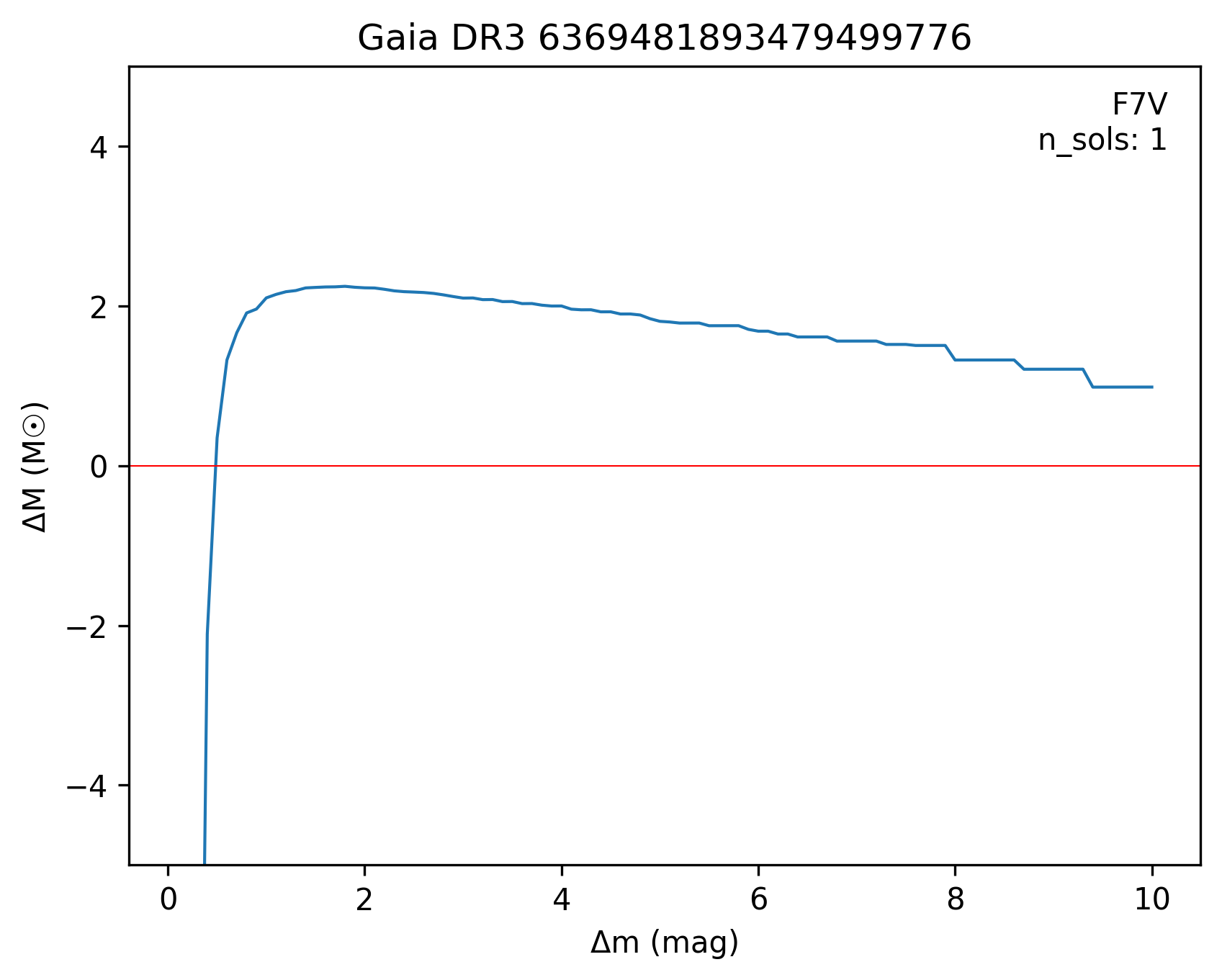}
    \end{subfigure}
\end{figure*}
\begin{figure*}[ht]
    \centering
    \begin{subfigure}[b]{0.45\textwidth}
    \includegraphics[width=\textwidth]{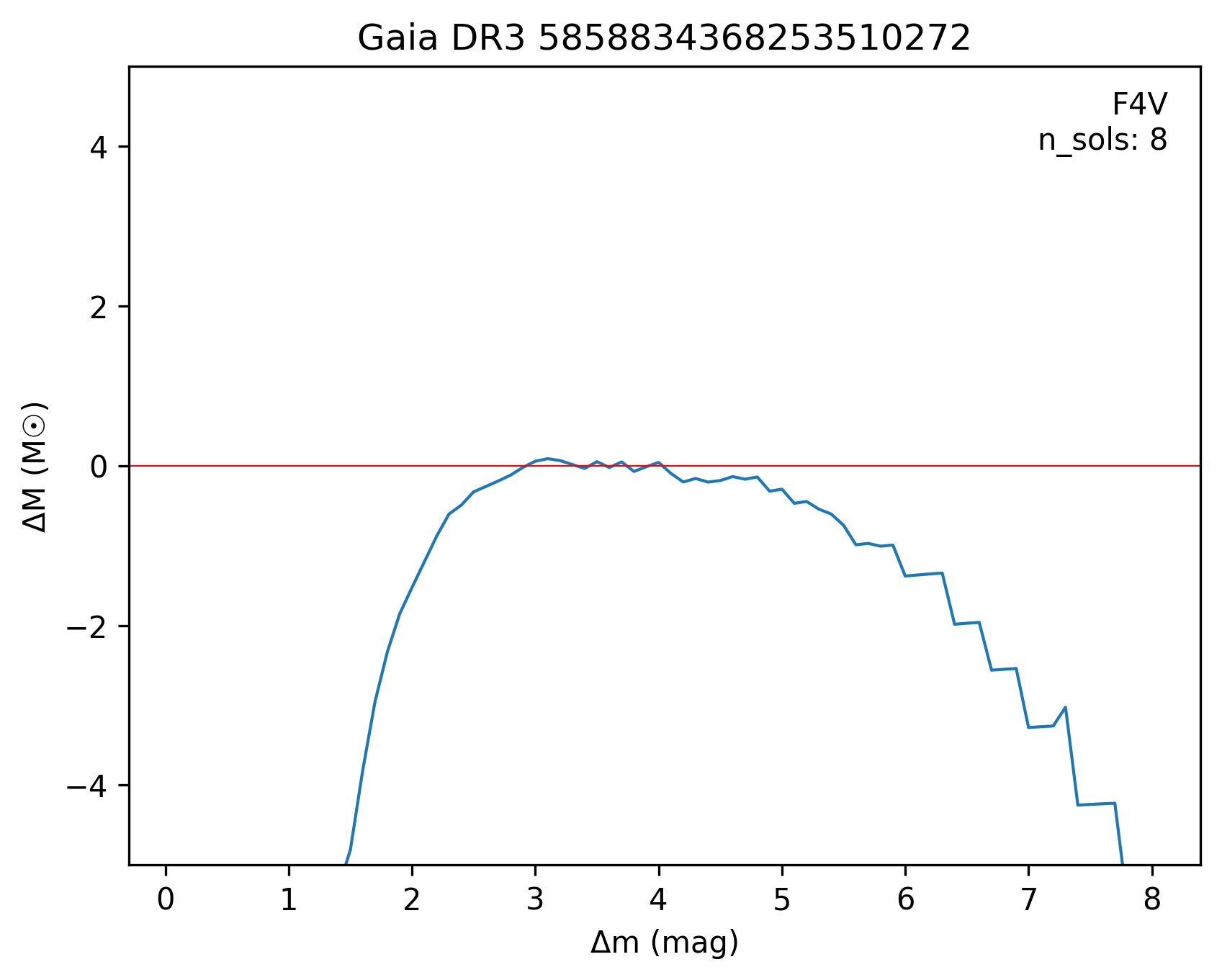}
    \end{subfigure}
     \hfill
    \begin{subfigure}[b]{0.45\textwidth}
    \includegraphics[width=\textwidth]{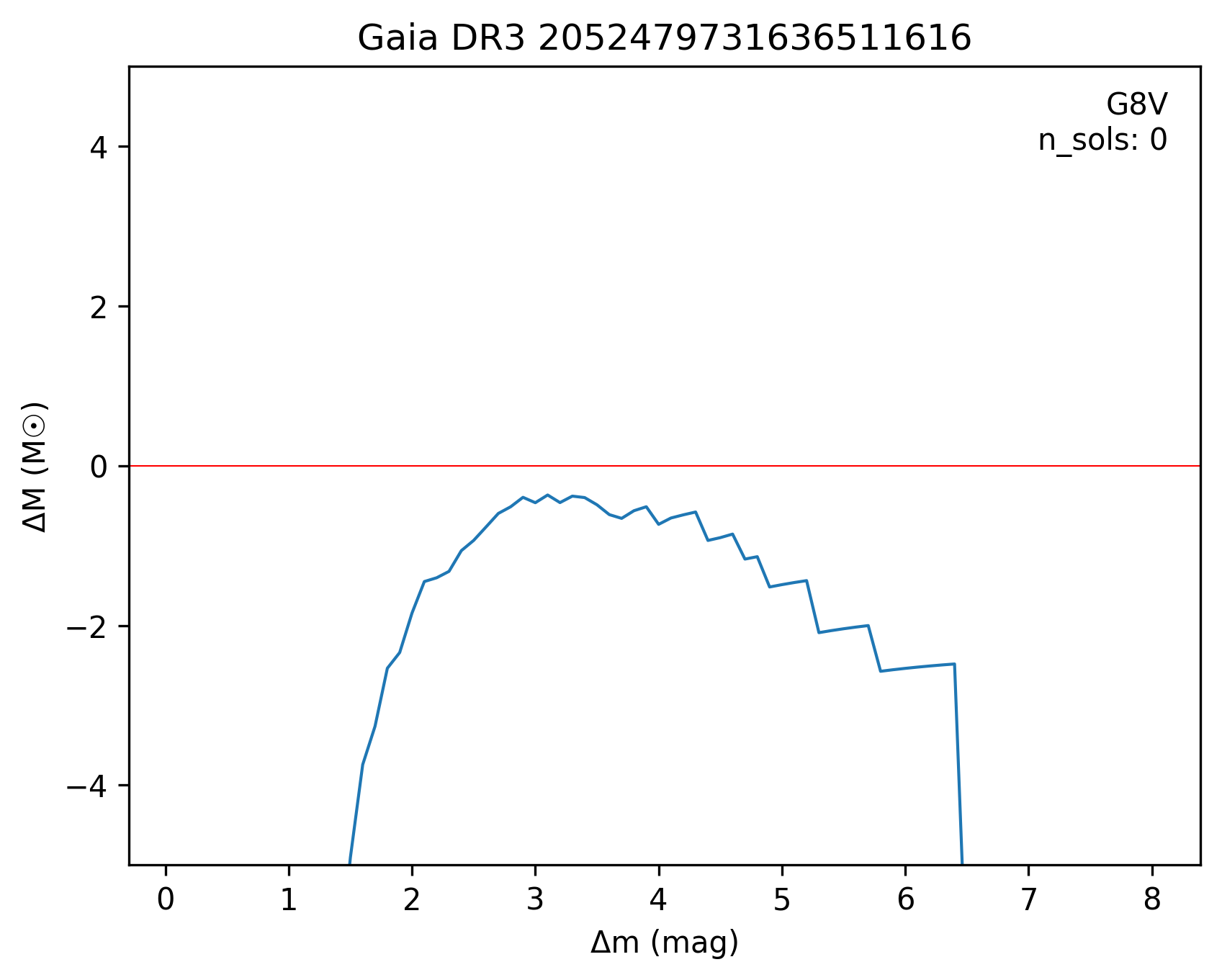}
    \end{subfigure}
     \caption{Graphical representation of $\Delta \mathcal{M}$ for a Gaia DR3 unresolved astrometric binaries, with 2 solutions (above, left), a single solution (above, right), 8 solutions (below, left) and without solution (below, right) for $\Delta m$.}
     \label{numericalplots}
\end{figure*}

As can be seen in Figure \ref{numericalplots} the function $\Delta \mathcal{M}$ shows an asymmetric parabolic shape. Considering its definition \eqref{deltamathcal}, that means that for lower values of $\Delta m$ the sum of orbital masses grows slower than the sum of spectroscopic ones, until they reach a minimum. Subsequently, the orbital term of the function becomes more important and $\Delta \mathcal{M}$ begins to decrease. Now, depending on the star, this function can cut the abscissa axis in different points. For instance, in figure \ref{numericalplots} (first row) we can see how $\Delta \mathcal{M}$ intersects with $\Delta \mathcal{M}=0$ in two points (that is, two possible solutions) and one point (but one more could be expected for larger $\Delta m$), 8 points (very close between them) and none, respectively.

We must note that, although the third subfigure shows a lot of solutions, they are due to the fact that the maximum of the parable is located very near of the line $\Delta \mathcal{M}=0$, so that instead of the two intersections between the line and an ideal parable, we get a higher number due to the uncertainties or because the chosen step is not small enough.

Regarding the no-solution case, corresponding to an asymmetric parable that does not touch the abscissa axis, it should be reminded that, even though we are evaluating $\Delta \mathcal{M}$ as a function of $\Delta m$, its values also depend of the physical ($Sp_T$) and dynamical ($\alpha, \pi'', P$) properties of the astrometric binary. 
In fact, if we look at the multiplicative factor $\gamma = \dfrac{\alpha^3}{\pi''^3 P^2}$ in the orbital term of $\Delta \mathcal{M}$, since $\gamma >0$ the larger $\gamma$ is, the more distant is $\Delta \mathcal{M}$ from the abscissa axis. Thereby, if $\Delta \mathcal{M}$ does not touch the zero line could indicate that the photocentric Gaia DR3 orbit is not right. Another possible reason is that the assumption taken by our model, that the photocentric orbit corresponds to a binary system, is incorrect: it belongs to a system with more companions (triple, quadruple, $\dots)$. For a sample of 796 Gaia DR3 astrometric binaries with main-sequence spectral type available in \textit{SIMBAD}, the mean and median number of solutions were two.

\subsection{Validation of the algorithm}

\begin{table*}[h!]
\footnotesize
\begin{center}
\caption{Comparison between the semi-major axis of 9 visual binary orbits found in the WDS-ORB6 and the results from the Gaia DR3 astrometric binaries computed by our algorithm}
\begin{tabular}{llcccc}
           &                                 & & WDS-ORB6        & & Our work       \\ \toprule
WDS & Gaia DR3 & $\Delta m\; (mag)$ & $a'' \pm \sigma_{a''}$ & $\Delta m\:(mag)$ & $a'' \pm \sigma_{a''}$ \\ \midrule
21424+3837 & 334156706263100416  & $2.89$       & $0.0255 \pm 0.0028$         & $1.45$        & $0.0282 \pm 0.0003$       \\ 
           &                     &            &                          & $6.35$        & $0.0268 \pm 0.0004$       \\ 
02572-2458 & 5076269164798852864 & $1.06$       & $0.0620 \pm 0.0020$           & $0.75$        & $0.0611 \pm 0.0004$       \\ 
           &                     &            &                          & $8.85$        & $0.0589 \pm 0.0004$       \\ 
04247+0442 & 3283823387685219328 & $1.00$          & $0.0113 \pm 0.0000$          & $2.15$        & $0.0115 \pm 0.0001$       \\ 
           &                     &            &                          & $4.85$        & $0.0108 \pm 0.0000$       \\ 
04375+1509 & 3309493720019304576 & Unknown          & $0.037$                    & $0.95$        & $0.0465 \pm 0.0006$       \\ 
           &                     &            &                          & $7.95$        & $0.0393 \pm 0.0007$       \\ 
09275-5806 & 5306416671004618240 & $0.68$       & $0.0335 \pm 0.0335$         & $0.65$        & $0.0349 \pm 0.0003$       \\ 
           &                     &            &                          & $9.45$        & $0.0289 \pm 0.0003$       \\ 
12313+5507 & 1571145907856592768 & Unknown          & $0.1023 \pm 0.0005$         & $1.85$        & $0.0939 \pm 0.0011$       \\ 
           &                     &            &                          & $4.85$        & $0.0878 \pm 0.0009$       \\ 
18040+0150 & 4468231641147900928 & $1.60$        & $0.0472 \pm 0.0016$         & $1.35$        & $0.0476 \pm 0.0004$       \\ 
           &                     &            &                          & $6.25$        & $0.0416 \pm 0.0006$       \\ 
12114-1647 & 3569106488558337792 & $1.62$       & $0.0253 \pm 0.0023$         & $0.85$        & $0.0261 \pm 0.0001$       \\ 
           &                     &            &                          & $8.75$        & $0.0233 \pm 0.0003$       \\ 
19380+3353 & 2047188847334279424 & $0.50$        & $0.0410 \pm 0.0009$       & $0.06$        & $0.0420 \pm 0.0450$ \\ \bottomrule
\end{tabular}

\label{tab:coincide}
\end{center}
\end{table*}

\begin{table*}[h!]
\footnotesize{
\begin{center}
\caption{Visual binary orbits found in the WDS-ORB6 for which our algorithm computed different results for the semi-major axis.}
\begin{tabular}{llcccc}
           &                                 & & WDS-ORB6        & & Our work       \\ \toprule
WDS        & Gaia DR3                         & $\Delta m\: (mag)$ & $a'' \pm \sigma_{a''}$ & $\Delta m\: (mag)$ & $a'' \pm \sigma_{a''}$ \\ \midrule
06047-4505 & 5567901976544151168    & $2.96$       & $6.0680 \pm  4.6020$        & $1.55$       & $0.0357 \pm 0.0003$      \\
           &                                 &            &                        & $6.25$       & $0.0310 \pm 0.0002$       \\
07204-5219 & 5492026740697659648   & $0.51$       & $600.0000$                    & $0.75$       & $0.0351 \pm 0.0002$      \\
           &                                 &            &                        & $9.05$       & $0.0274 \pm 0.0002$      \\
08031-0625 & 3067074530201582336   & $0.27$       & $2.3222 \pm 0.9850$     & $0.55$       & $0.0045 \pm 0.0000$         \\
           &                                 &            &                        & $9.05$       & $0.0034 \pm 0.0000$         \\
08165+7930 & 1139059897091616512   & $0.24$       & $15.0000 \pm 2.0000$                & $0.75$       & $0.0022 \pm 0.0000$         \\
           &                                 &            &                        & $8.95$       & $0.0018 \pm 0.0000$         \\
11214-2027 & 3545469496823737856   & $2.39$ & $6.5328 \pm 1.9456$  & $0.55$ & $0.4998 \pm 0.0552$ \\
           &                                 &            &                        & $7.75$       & $0.3624 \pm 0.0493$      \\
12115+5325 & 1572914025633785728  & $0.16$ & $19.6680 \pm 9.1636$ & $1.15$ & $0.0608 \pm 0.0003$ \\
           &                                 &            &                        & $6.65$       & $0.0550 \pm 0.0003$       \\
13132-0501 & 3635663363361305088   & $2.6$        & $0.1630$                  & $0.95$       & $0.0426 \pm 0.0010$       \\
           &                                 &            &                        & $7.65$       & $0.0360 \pm 0.0009$       \\
18520-5418 & 6650457382678224640   & $2.89$       & $0.1040 \pm 0.0020$         & $0.45$       & $0.0429 \pm 0.0002$      \\
           &                                 &            &                        & $9.85$       & $0.0301 \pm 0.0002$      \\
19581-4808 & 6670759761800880768   & $2.17$       & $0.2364 \pm 0.0140$       & $0.75$       & $0.0485 \pm 0.0024$     \\
           &                                 &            &                        & $8.55$       & $0.0384 \pm 0.0020$       \\
23347+3748 & 1918953867019478144  & $0.3$        & $0.0690 \pm 0.0060$         & $0.45$       & $0.0270 \pm 0.0003$       \\
           &                                 &            &                        & $0.45$       & $0.0268 \pm 0.0003$      \\
05542-2909 & 2903665833633799040   & $1.84$       & $0.1117 \pm 0.0014$       & $1.15$       & $0.0737 \pm 0.0048$      \\
           &                                 &            &                        & $9.05$       & $0.0724 \pm 0.0049$      \\
15390-5742 & 5882581895219921024   & $2.9$        & $0.0797 \pm 0.0021$       & $0.55$       & $0.0569 \pm 0.0020$       \\
           &                                 &            &                        & $9.75$       & $0.0479 \pm 0.0018$  \\ \bottomrule
\end{tabular}

\label{tab:nocoincide}
\end{center}}
\end{table*}

In order to get some insights on the reliability of the solutions computed by our algorithm, we have cross-matched the orbital solutions for astrometric orbits given by the Gaia DR3 NSS table (namely, \texttt{Orbital}, \texttt{OrbitalTargetedSearch}, \texttt{OrbitalTargetedSearchValidated} and \texttt{AstroSpectroSB1}) with the orbits contained in the \textit{Sixth Catalog of Orbits of Visual Binary Stars} (WDS-ORB6). By doing that, we have constrain a set of 26 Gaia DR3 astrometric orbits that have a resolved relative orbit in the WDS-ORB6. Subsequently, by applying our algorithm to them, we have seen that 21 belong to the main sequence, attending both to the $\log(g)$ and to the difference between the calibrated and the Gaia-computed effective temperature ($<500K$). 

The comparison between the relative semi-major axes available in the WDS-ORB6 and those obtained by using our algorithm is shown in tables \ref{tab:coincide} and \ref{tab:nocoincide}.
In the first one, we can see that the most probables $a''$ obtained by the algorithm differs, in average, only a $3.7\%$ for the most similar solutions, and a $10.6\%$ if we take only the worst ones. On the other hand, in table \ref{tab:nocoincide} we can see the other 12 solutions, showing a difference of even several orders of magnitude in comparison with those cross-matched from WDS-ORB6. However, as we discuss below, some of these differences are very well characterized. Let's see each case individually.

\subsubsection{WDS 06047-4505}
The semi-major axis of this binary, calculated by \citet{izm2019}, is two orders of magnitude larger than those computed with our algorithm. This may indicate that this binary is actually a system of three or more companions, so that the Gaia DR3 astrometric orbit corresponds to a subcomponent of that stellar system. Indeed, the WDS 06047-4505 (also known as HIP 28790) is a quintuple system, where the semi-major axis, shown in table \ref{tab:nocoincide}, correspond to the orbit of two binary subsystems, A and B; while another companion (C) is located at more that $3'$ \citep{tok2016}.

Thus, the astrometric orbit obtained by Gaia may correspond to the subsystem HIP 28790 Aa-Ab or HIP 28790 Ba-Bb. Moreover, the orbital period ($P=221.19 \:d$) of the Gaia's orbit is in high consonance with that of the first subsystem ($P=221.4 \: d$) in opposition with the second one ($P=13.23 \: d$) \citep{tok2016}, much shorter. Therefore, we can conclude that our algorithm provides the relative semi-major axis for the stellar subsystem HIP 28790 Aa-Ab.

\subsubsection{WDS 07204-5219}
In this case, the ORB6 shows a huge semi-major axis of $10'$ and $P= 1000\:Gyr$ that, unsurprisingly, is not likely to be correct \citep{letchford2018}. However, it is known that WDS 07204-5219 is a hierarchical quintuple system formed by a binary star (HD 57583A) orbiting around a triple one (HD 57583A), being its main orbital period, $P= 122 \: d$ \citep{saar1990, desidera2006}. Thereafter, the solutions provided by the algorithm are expected to match with those of one of the subcomponents, although further ground-based measurements are needed to confirm it.

\subsubsection{WDS 08031-0625}
The orbital period shown in the WDS-ORB6 for this binary is $P=509.87 \:d$ \citep{izm2019}, while the obtained astrometrically from Gaia is $P=43.63 \:d$, therefore it may corresponds to a subcomponent yet unknown. This would justify the difference in three orders of magnitude between the semi-major axis of the already known orbit and the obtained by our algorithm. However, more observations are needed in order to confirm that orbit.

\subsubsection{WDS 08165+7930}
Here we have a triple system (ADS 6646), where the orbital period of ADS 6646 A-B is registered as $P=200 \: centuries$ in WDS-ORB6 \citep{kis2009}, and that of ADS 6646 Aa-Ab is $P=4.88 \:d$ \citep{tok1997}. Since the period of the Gaia's astrometric orbit is $P=14.33 \:d$, it is clear that it corresponds neither to the first nor to the second orbit. Moreover, if we zoom out the search radius in the cross-match up to $15'$, one more source, catalogued as single star, appears in the Gaia Archive, which it is arguably the ``B'' companion. Consequently, we have a discrepance between the astrometric ADS 6646 Aa-Ab orbit period and their spectroscopic one. It should be noticed that the SB1 orbit of that binary was measured in 1997, so that new measurements must be carried out in order to study the reason of this discrepance between Gaia's and Tokovinin's results. A possible solution is that the triple system is not hierarchical, so that the astrometric orbit can not be transformed into a binary one by means of our methodology.

\subsubsection{WDS 11214-2027}
As can be seen in table \ref{tab:nocoincide}, the semi-major axis from ORB6 is about an order of magnitude larger than those obtained with the algorithm, being its orbital period ($P=1028.09 \:yr$) \citep{izm2019}, very different from the Gaia's one ($P=13.36 \:d$), pointing out that they may belong to different orbits within the same system. However, this star is known in the literature as a binary, since only two components have been discovered. This seems surprising because of the large algorithmic semi-major axis $a''=0.4998''$: However, if the second solution ($a''=0.3624''$) is correct, even with such a large separation it would be difficult to resolve that subsystem with a magnitude difference of $\Delta m= 7.75\:mag$.

\subsubsection{WDS 12115+5325}
The orbital period of this system, as it is shown in the WDS-ORB, is $P=2434.57 \:yr$ \citep{izm2019}, while that detected by Gaia is $P=1.77 \:yr$. This huge difference between periods justify the proportional discrepance in semi-major axes, indicating that our orbit may correspond to a subcomponent yet unresolved of the WDS 12115+5325 system.

\subsubsection{WDS 13132-0501}
This system is catalogued as binary, with an orbital period from the spectroscopic orbit of $P=17 \:yr$ \citep{tok2014}. Neverthless, the orbital period obtained by Gaia is $P=1.76 \:yr$, therefore the corresponding orbit belongs to a subcomponent of the system.

\subsubsection{WDS 18520-5418}
In this triple system the primary component A is a binary with an orbital period of $P=11.25 \:yr$ and $a''=0.104''$ \citep{circ204}. However, the orbital period measured by Gaia is $P=2.36 \:yr$ indicating that it may correspond to a different component in that system. Since the component B is located at $146''$ \citep{toklep2012}, the data points out that either Gaia's or Tokovinin's orbit is not good or that one of the subcomponents, Aa or Ab, is actually a binary itself.

\subsubsection{WDS 19581-4808} 
The binary WDS 19581-4808, or HIP 98274, has a definitive orbit with an orbital period of $P=35.28 \:yr$ \citep{tok2016} so that Gaia's astrometric orbit with a period of $3.08 \:yr$ corresponds to a subcompanion not yet discovered.

\subsubsection{WDS 23347+3748}
The semi-major axis of WDS 23347+3748, calculated for the first time by \citet{horch2015}, shows a discrepance with that computed by using our algorithm, as well as the ground-measured orbital period $P= 3.64 \:yr$ which is very different from that observed by Gaia ($P=0.95 \:yr$). However, a more recent spectroscopic orbit was obtained for this system by \citet{kiefer2018}, showing a period of $P=~0.95 \:yr$ and an eccentricity $e=0.435$ which matches with that of the Gaia astrometric orbit ($e=0.441$). In addition to this, that spectroscopic orbit, available at the \textit{Ninth Catalogue of Spectroscopic Binary Orbits} (\url{https://sb9.astro.ulb.ac.be/}), includes the individual semi-major axis of each component multiplied by the $\sin$ of the inclination, $I$: $a_1''\sin(I) = 8.77468\cdot 10^{7}\:km$ and $a_2''\sin(I) = 9.42625\cdot 10^{7}\:km$. Therefore, just using the inclination computed in Gaia DR3, the Gaia's parallax for this object and the fact that $a''=a_1''+a_2''$, we have: $a''=0.0250''$, that is in consonance with those computed by our algorithm: $a''=0.0270''$ or $0.0268''$. On consequence, we can conclude that the orbit available in the WDS-ORB6 is not correct, and that one of those calculated through our algorithm are more likely to be right.

\subsubsection{WDS 05542-2909}
The orbital period of WDS 05542-2909 that appears in the ORB6 is $P=10.07 \: yr$, whereas the computed by Gaia is $P=4.98 \:yr$. That difference between the orbital periods shows that the astrometric orbit detected by Gaia may correspond to a subcomponent, although specific observations should be carried out in order to confirm it.

\subsubsection{WDS 15390-5742}
The star WDS 15390-5742 (or HD 139084) is a hierarchical triple system in the $\beta$ Pictoris moving group, with a main orbit with $a''= 10''$ and a subsystem in the primary component of $P=4.58 \: yr$ \citep{nielsen2016}, that does not match the period of the Gaia-detected astrometric orbit, of $P=2.46 \: yr$. Like the previous cases, we should expect that the astrometric orbit corresponds to a subcomponent of the triple system.

\subsection{Application on the  Gaia DR3 astrometric orbits: the ESMORGA catalog}

By applying the criteria shown in subsection \ref{initialsample}, we have constrained an initial sample of 86586 astrometric binaries. By applying the algorithm to all of them, we have obtained solutions for a total of 60 383 systems, and dumped them into a catalog, hereafter ESMORGA (Ephemeris, Stellar Masses and relative ORbits from GAia), that can be downloaded from the Ramón María Aller Astronomical Observatory website: in HTML for simple queries (\url{https://www.usc.es/astro/esmorga.html}) or in CSV (\url{https://www.usc.es/astro/esmorga.txt}) for deeper analysis. Here, we present a description of the catalog, column by column.

\subsubsection{Description of the catalog}

Each binary star corresponds to an even row, while the uncertainties are shown in the row below. If two or more possible solutions for each star are provided, they appear in consecutive rows.
The columns with the information given for each source are organized as follows:

\begin{itemize}[leftmargin=0.7cm]
    \item[1.] Gaia DR3 source identifier, \textit{gaia\_id}
    \item[2.] Solution type: \texttt{Orbital}, \texttt{Orbital*}\footnote{\texttt{Orbital*}=\texttt{OrbitalTargetedSearch}} and \texttt{Orbital**}\footnote{\texttt{Orbital**}=\texttt{OrbitalAlternative}} for astrometric orbits from differente datasets and processing models, and \texttt{AstroSpectroSB1} for those with both astrometric and spectroscopic solutions.
    \item[3.] Right Ascensión for J2016.0, \textit{RA} ($^{\circ}$)
    \item[4.] Declination for J2016.0, \textit{Dec} ($^{\circ}$)
    \item[5.] Parallax, $\pi''$ ($mas$)
    \item[6.] Apparent composite magnitude in Johnson-Cousins $V$-band, $m_V$ ($mag$)
    \item[7.] Absolute composite magnitude in Johnson-Cousins $V$-band, $M_V$ ($mag$)
    \item[8.] Composite spectrum in MK designation, $Sp_T$
    \item[9.] Magnitude difference, $\Delta m$ ($mag$)
    \item[10.] Absolute magnitude for the primary in Johnson-Cousins $V$-band, $M_{VA}$ ($mag$)
    \item[11.] Absolute magnitude for the secondary in Johnson-Cousins $V$-band, $M_{VB}$ ($mag$)
    \item[12.] Spectral type of the primary in MK designation, $Sp_A$
    \item[13.] Spectral type of the secondary in MK designation, $Sp_B$
    \item[14.] Mass of the primary, $\mathcal{M}_A$ ($\mathcal{M}_{\odot}$)
    \item[15.] Mass of the secondary, $\mathcal{M}_B$ ($\mathcal{M}_{\odot}$)
    \item[16.] Effective temperature of the primary, $T_{effA}$ ($K$)
    \item[16.] Effective temperature of the secondary, $T_{effB}$ ($K$)
    \item[17.] Period, $P$ ($yr$)
    \item[18.] Time of periastron, $T$ (\textit{Fractional decimal year})
    \item[19.] Eccentricity, $e$
    \item[20.] Semi-major axis, $a''$ ($''$)
    \item[21.] Inclination, $I$ ($^{\circ}$)
    \item[22.] Longitude of the node, $\Omega$ ($^{\circ}$)
    \item[23.] Argument of the periastron, $\omega$ ($^{\circ}$)
    \item[24.] Maximum angular separation of the apparent orbit, $\rho''_{max}$ ($''$)
    \item[25.] Date of $\rho''_{max}$ (\textit{Fractional decimal year})
    \item[26.] Minimum angular separation of the apparent orbit, $\rho''_{min}$ ($''$)
    \item[27.] Date of $\rho''_{min}$ (\textit{Fractional decimal year})
    \item[28.] Grade of consonance with \small{\texttt{gaiadr3.binary\_masses}}
    \begin{itemize}
        \item \texttt{grade} = 1 if  $\dfrac{|\mathcal{M}_A - \texttt{m1}|}{\texttt{m1}} \leq 0.1$
        \item \texttt{grade} = 2 if  $0.1 < \dfrac{|\mathcal{M}_A - \texttt{m1}|}{\texttt{m1}} \leq 0.2$
        \item \texttt{grade} = 3 if  $0.2 < \dfrac{|\mathcal{M}_A - \texttt{m1}|}{\texttt{m1}} \leq 0.3$
        \item \texttt{grade} = 4 if  $0.3 < \dfrac{|\mathcal{M}_A - \texttt{m1}|}{\texttt{m1}} \leq 0.4$
        \item \texttt{grade} = 5 if  $0.4 < \dfrac{|\mathcal{M}_A - \texttt{m1}|}{\texttt{m1}} \leq 0.5$
    \end{itemize}
    \normalsize
    In the case of astrospectroscopic orbits, we take into consideration both masses, by using as reference in the above comparison the following term instead of operating only with the primary one:
    \begin{equation}
    max\left\{\dfrac{|\mathcal{M}_A - \texttt{m1}|}{\texttt{m1}}, \dfrac{|\mathcal{M}_B - \texttt{m2}|}{\texttt{m2}}\right\}
    \end{equation}
         
\end{itemize}

\subsubsection{Catalog statistics}

\begin{figure}[h!]
    \centering
    \includegraphics[width=0.45\textwidth, trim={0 0.6cm 0 0.9cm},clip]{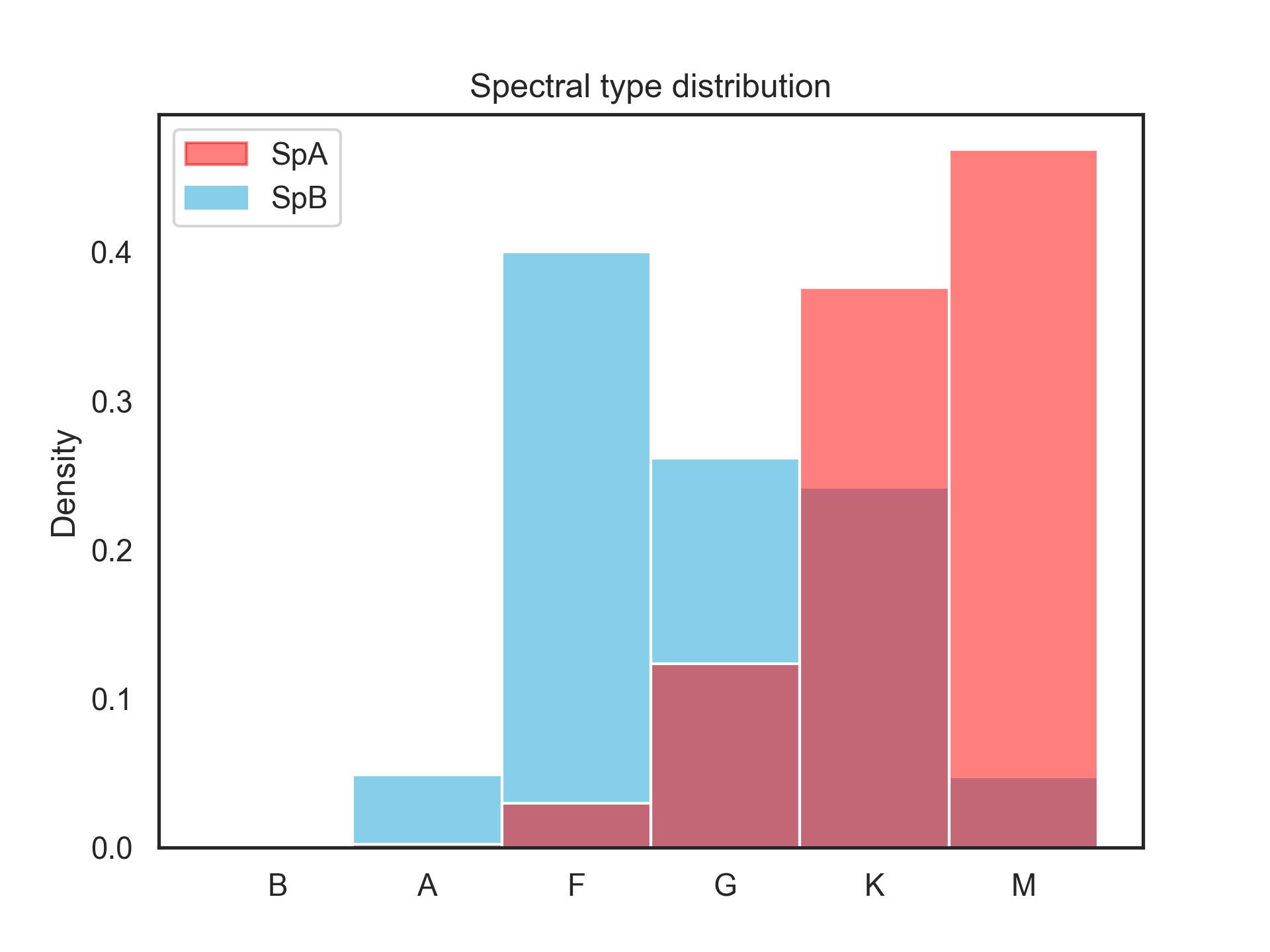}
    \caption{Spectral type distribution of the 121 028 stellar companions contained in ESMORGA}
    \label{sp_distribution}
\end{figure}

Among the 60383 binaries contained in ESMORGA, there are $43$ ($0.04 \%$) B-types, $3782$ ($3.54 \%$) A-types, $35 857$ ($33.56 \%$) F-types, $33014$ ($30.90\%$) G-types, $28617$ ($26.78 \%$) K-types and $5545$ ($5.19\%$) M-types. That is in regards to the composite spectrum. If we now look at the disentangled spectrum of each companion, shown in Figure \ref{sp_distribution} we can also see that, while three-quarters of the primary components are F and G-types, cooler K and M-type primaries are not uncommon. Second of all, the most frequent secondaries are, by far ($~85 \%$), red and orange dwarfs. 

Regarding the other astrophysical properties of the Gaia astrometric binaries that can be now studied through this new catalog, their geometrical distribution and, in particular, the separations between their companions, is arguably the most straightforward one.

\begin{figure}[ht]
    \centering
    \includegraphics[width=0.49\textwidth, trim={0.5cm 0 1.5cm 0.85cm},clip]{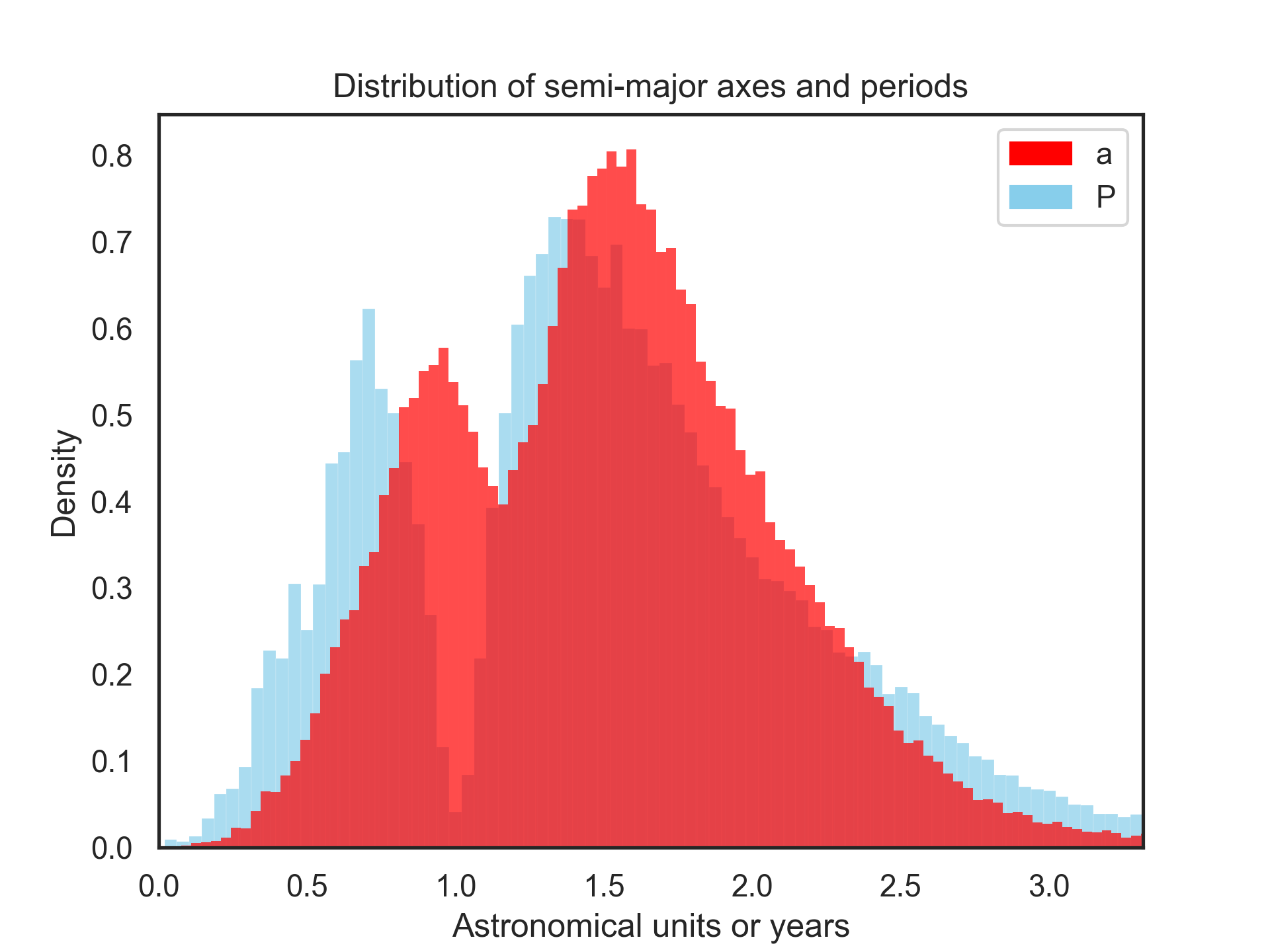}
    \caption{Density histogram of $a$ and $P$ for 60 514 binaries}
    \label{a+P histogram}
\end{figure}

In Figure \ref{a+P histogram} the density histogram of the semi-major axes in astronomical units is shown in red. The sample reaches the $99^{th}$ percentile to enhance data visualization. As can be seen, Gaia astrometric binaries are very close binaries, separations within the Solar System scale, having the $~50\%$ binaries a semi-major axis of $~2\:au$. Moreover, there is a deep decay around $1\:au$, that if we compare it with the histogram of the periods (blue), it is clear that it matches with Gaia's observational bias for binary orbits of $~1\:yr$, due to the coupling of the orbital and parallactic motions, originated by the spacecraft’s
movement \citep{halbwachs, hiddentreasure}.

\begin{figure*}[ht!]
\begin{minipage}{0.46\textwidth}
    \includegraphics[trim={0 0 0 0}, width=\textwidth, clip]{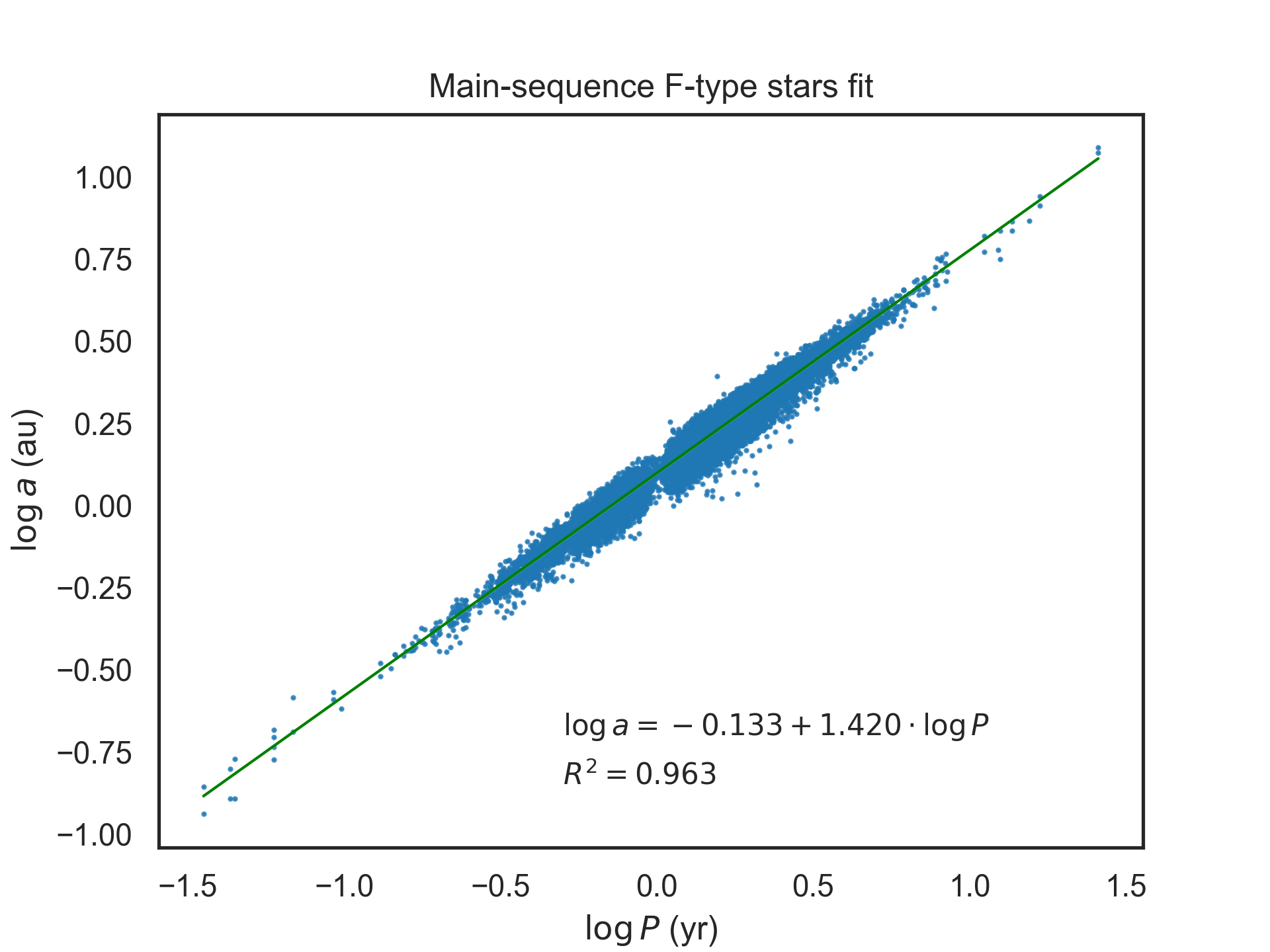}
    \captionof{figure}{$\log{a}-\log{P}$ diagram for our algorithmic orbital solutions on main-sequence F-type Gaia DR3 astrometric binaries}
    \label{F-typefit}
\end{minipage}
\hspace{1cm}
\begin{minipage}{0.46\textwidth}
    \includegraphics[width=\textwidth, trim={0 0 0 0}, clip]{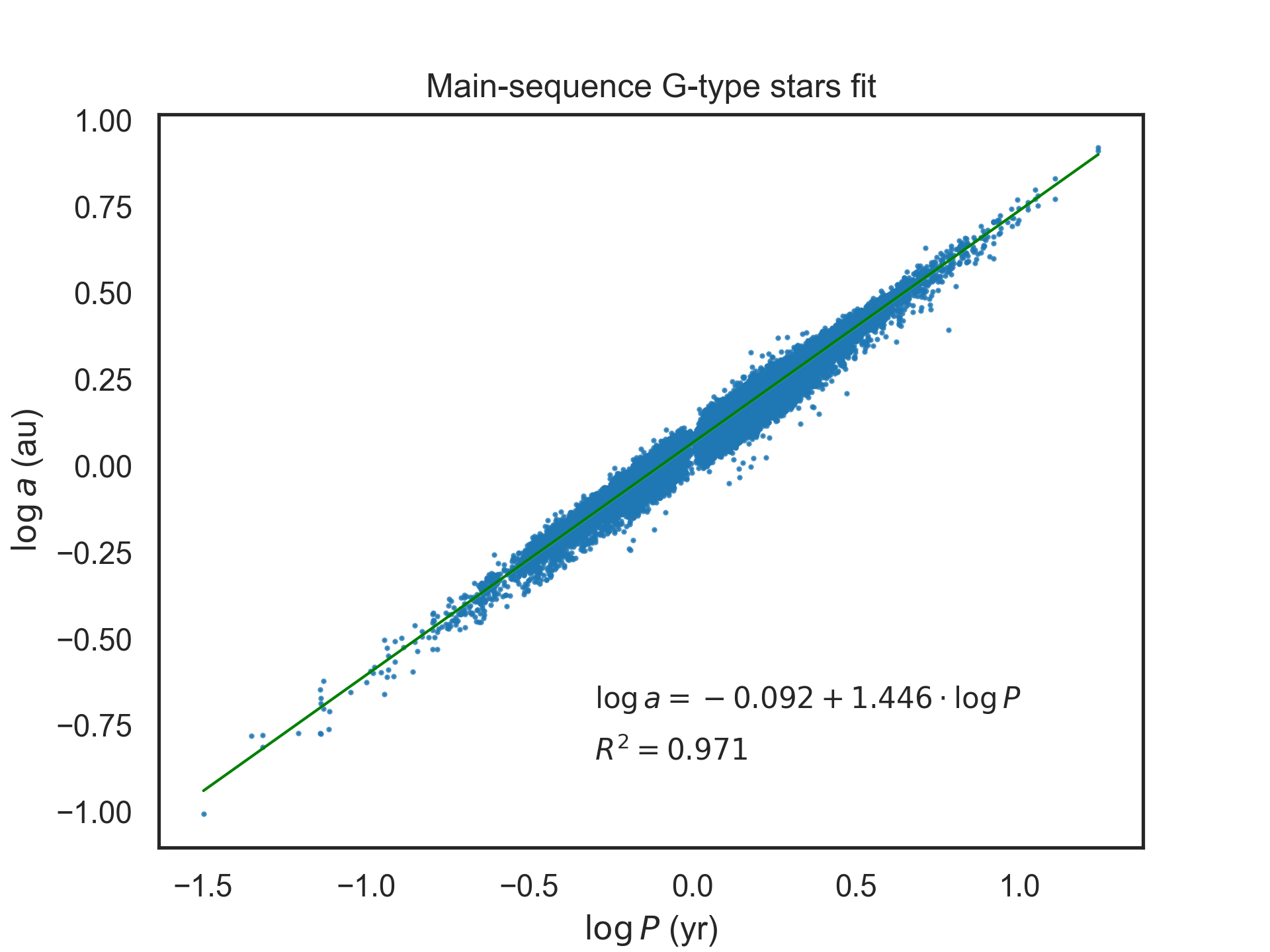}
    \captionof{figure}{$\log{a}-\log{P}$ diagram for our algorithmic orbital solutions on main-sequence G-type Gaia DR3 astrometric binaries}
    \label{G-typefit}
\end{minipage}
\begin{minipage}{0.46\textwidth}
    \includegraphics[width=\textwidth, trim={0 0 0 0}, clip]{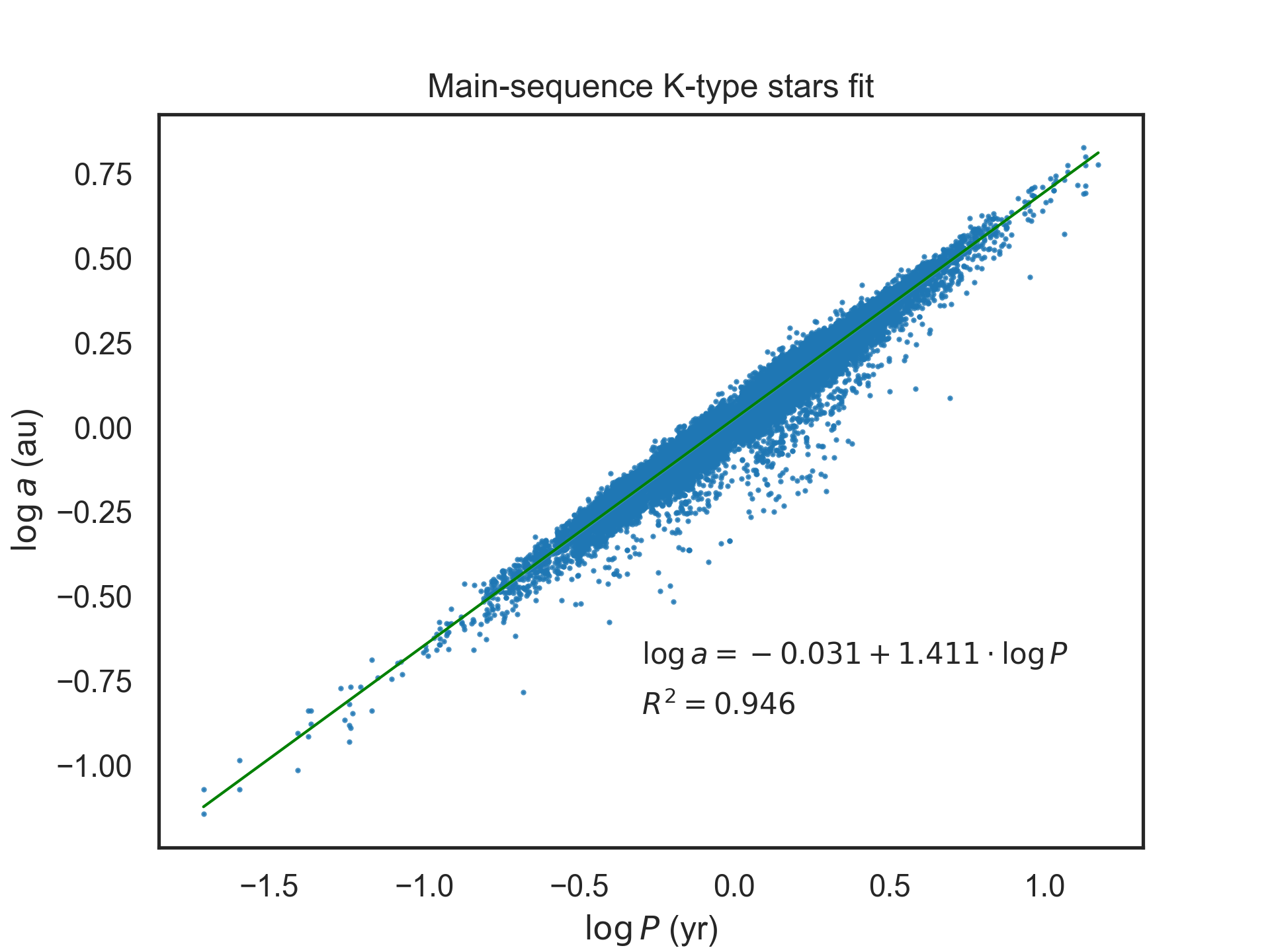}
    \captionof{figure}{$\log{a}-\log{P}$ diagram for our algorithmic orbital solutions on main-sequence K-type Gaia DR3 astrometric binaries}
    \label{K-typefit}
\end{minipage}
\hspace{1cm}
\begin{minipage}{0.46\textwidth}
    \includegraphics[width=\textwidth, trim={0 0 0 0}, clip]{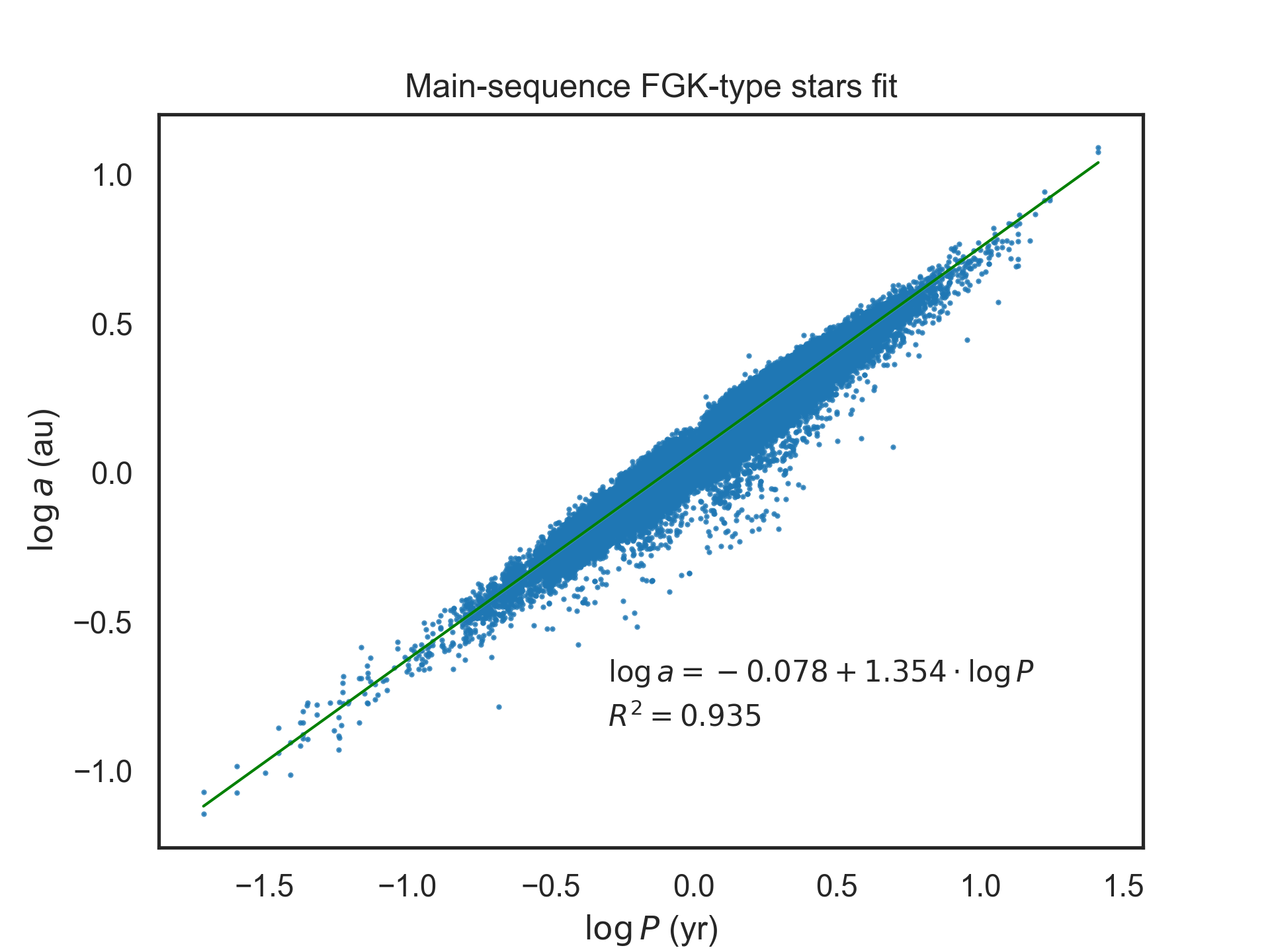}
    \captionof{figure}{$\log{a}-\log{P}$ diagram for our algorithmic orbital solutions on main-sequence FGK-types Gaia DR3 astrometric binaries}
    \label{FGK-typefit}
\end{minipage}
\begin{minipage}{0.46\textwidth}
\footnotesize
\captionof{table}{\label{fitcoefF} Parameters of the main-sequence F-type $\log{P} - \log{a}$ fit in figure \ref{F-typefit}, expressed in the form \eqref{polynomialpowers}}
\begin{center}
\begin{tabular}{p{0.2\textwidth}p{0.2\textwidth}p{0.2\textwidth}p{0.2\textwidth}}
\midrule
Parameter & Value  & Std error & p-value          \\ \toprule
$a_0$     & -0.1329 & 0.000     & \textless{}0.001 \\
$a_1$     &  1.4202 & 0.001     & \textless{}0.001 \\
\textit{RSE}       & 0.042 &           &                  \\
$R^2$     & 0.963  &           &  
        \\ \bottomrule
\multicolumn{4}{l}{\textit{RSE: Residual standard error (on 35 855 degrees of freedom)}}
\end{tabular}
\end{center}
\end{minipage}
\hspace{1cm}
\begin{minipage}{0.46\textwidth}
\footnotesize
\captionof{table}{\label{fitcoefG} Parameters of the main-sequence G-type $\log{P} - \log{a}$ fit in figure \ref{G-typefit}, expressed in the form \eqref{polynomialpowers}}
\begin{center}
\begin{tabular}{p{0.2\textwidth}p{0.2\textwidth}p{0.2\textwidth}p{0.2\textwidth}}
\midrule
Parameter & Value  & Std error & p-value          \\ \toprule
$a_0$     & -0.0917 & 0.000     & \textless{}0.001 \\
$a_1$     &   1.4456 & 0.001     & \textless{}0.001 \\
\textit{RSE}       & 0.042 &           &                  \\
$R^2$     & 0.971  &           &  
        \\ \bottomrule
\multicolumn{4}{l}{\textit{RSE: Residual standard error (on 33 012 degrees of freedom)}}
\end{tabular}
\end{center}
\end{minipage}
\begin{minipage}{0.46\textwidth}
\footnotesize
\captionof{table}{\label{fitcoefK} Parameters of the main-sequence K-type $\log{P} - \log{a}$ fit in figure \ref{K-typefit}, expressed in the form \eqref{polynomialpowers}}
\begin{center}
\begin{tabular}{p{0.2\textwidth}p{0.2\textwidth}p{0.2\textwidth}p{0.2\textwidth}}
\midrule
Parameter & Value  & Std error & p-value          \\ \toprule
$a_0$     & -0.0305 & 0.000     & \textless{}0.001 \\
$a_1$     &  1.4114 & 0.002     & \textless{}0.001 \\
\textit{RSE}       & 0.062 &           &                  \\
$R^2$     & 0.946  &           &  
        \\ \bottomrule
\multicolumn{4}{l}{\textit{RSE: Residual standard error (on 28 615 degrees of freedom)}}
\end{tabular}
\end{center}
\end{minipage}
\hspace{1cm}
\begin{minipage}{0.46\textwidth}
\footnotesize
\captionof{table}{\label{fitcoefFGK} Parameters of the main-sequence FGK-types $\log{P} - \log{a}$ fit in figure \ref{FGK-typefit} (left), expressed in the form \eqref{polynomialpowers}}
\begin{center}
\begin{tabular}{p{0.2\textwidth}p{0.2\textwidth}p{0.2\textwidth}p{0.2\textwidth}}
\midrule
Parameter & Value  & Std error & p-value          \\ \toprule
$a_0$     & -0.0776 & 0.000     & \textless{}0.001 \\
$a_1$     & 1.3538 & 0.001     & \textless{}0.001 \\
\textit{RSE}       & 0.063 &           &                  \\
$R^2$     & 0.935  &           &  
        \\ \bottomrule
\multicolumn{4}{l}{\textit{RSE: Residual standard error (on 97 486 degrees of freedom)}}
\end{tabular}
\end{center}
\end{minipage}
\end{figure*}

Another informative relation is the oldie but goldie $a-P$ relation in which we can deepen thanks to our working sample. Indeed, if we graph the scatter plot of $\log{a} - \log{P}$ for F, G, and K main-sequence binaries, it could be seen that they follow a linear and homoscedastic distribution so that we can establish a linear regression model that fits very well with observable data, as shown in figures \ref{F-typefit}, \ref{G-typefit}, \ref{K-typefit} and \ref{FGK-typefit} for late-type main-sequence stars.

Therefore, the semi-major axis in astronomical units can be expressed, for late-type main-sequence close binaries, as a function of the period by means of polynomial powers of ten, that is:
\begin{equation}\label{polynomialpowers}
    a(P) = 10^{a_0 + a_1\log{P}}
\end{equation}

The fitting coefficients, $a_0$ and $a_1$, as well as some indicators about the quality of these fits, are shown for each calibration in tables \ref{fitcoefF}, \ref{fitcoefG}, \ref{fitcoefK} and \ref{fitcoefFGK}.
For M-dwarfs and early types (B and A) the scatter plots do not show enough linearity to perform any linear regression fit, and neither a polinomial one, due to the high observed heteroscedasticity.
Notwithstanding that, and due to the goodness of the other fits, we suggest them as a useful tool for estimating, in first approximation, the semi-major axis $a$ (in distance units) for any Gaia DR3 astrometric binary of those spectral classes, just by using its orbital period. Subsequently, it allows to compute an estimation of the total mass of each unresolved system as well as its ephemeris, as described in subsection \ref{algorithm}.

In the case of astrospectroscopic orbits, it should be noticed that, since we have $I$ from the astrometric orbit, $a_1$ is straightforwardly obtained and, by means of the following relation, 
\begin{equation}
    \dfrac{a}{\mathcal{M}_{SpA} + \mathcal{M}_{SpB}} = \dfrac{a_1}{\mathcal{M}_{SpB}} = \dfrac{a_2}{\mathcal{M}_{SpA}},
\end{equation}
the semi-major axis of the secondary component around the center of mass,
$a_2 = a_1 \frac{\mathcal{M}_{SpA}}{\mathcal{M}_{SpB}}$, is known.

As a consequence, we can derive an alternative relative semi-major axis $\hat{a} = a_1 + a_2$. Now, by dividing the relative semi-major axis obtained by the algorithm (in arc seconds) by $\hat{a}$ (in $au$), an alternative value $\hat{\pi}'' = a''/\hat{a}$ computed with new data is derived for the parallax (hereafter, the astrospectroscopic parallax).

Thus, it is clear that, if the relative orbit aligns with the Gaia spectroscopic one, then $\pi''\approx \hat{\pi}''$. Indeed, from the comparison of the 8583 astrospectroscopic parallaxes with those of Gaia, we have found a mean absolute discrepance of $0.43\:mas$ ($std=0.52\: mas$) that corresponds to a mean relative difference of $8.93\%$ between them. Therefore, we suggest that as a double validation on the quality of the procedure with astrospectroscopic solutions.
  
\begin{figure}[ht]
    \centering
    \includegraphics[trim={0.5cm 0 1.2cm 0.85cm}, width=0.5\textwidth, clip]{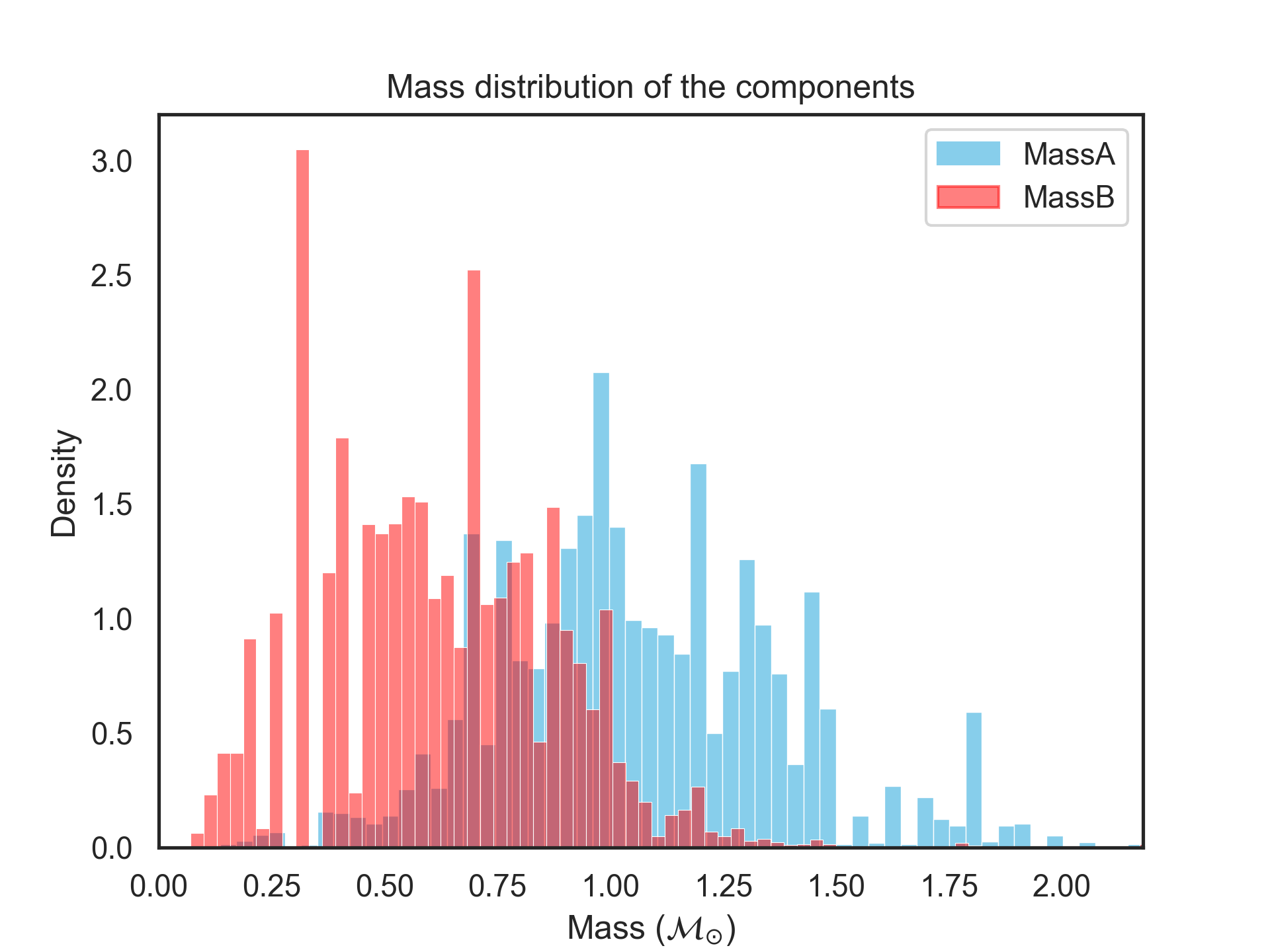}
    \caption{Mass distribution on the primary (blue) and secondary (red) components.}
    \label{massdistribution}
\end{figure}

Finally, we present a histogram in Figure \ref{massdistribution} with the distribution of the masses for primary and secondary components, using the methodology of this work. Here we have cut the sample at the $99.9^{th}$ percentile, although the maximum mass obtained was $3.69 \mathcal{M}_{\odot}$ for a primary and $2.99 \mathcal{M}_{\odot}$ for a secondary. The minimum masses of the catalog correspond to a primary of $0.10 \mathcal{M}_{\odot}$ and a secondary of $~0.07 \mathcal{M}_{\odot}$, both in the limit of the calibrations. 

\section{Conclusions}

We have developed and implemented an analytic algorithm to calculate the semi-major axis for the relative orbit of main-sequence and subgiant Gaia DR3 astrometric binaries, providing the most probable value for $a$ in the case of an astrometric solution and the uniquely correct solution for the astrospectroscopic ones. Through our procedure, several decisive astrophysical parameters such as the spectral types, stellar masses, and effective temperatures are derived for each component of those unresolved systems. Moreover, the ephemeris for the apparent orbit of those systems is obtained, to know the angular separations of their companions to resolve them, and thus to confirm it. The results for 60 383 binary systems using that methodology are presented in the ESMORGA catalog described in this work.

We have validated our algorithm by comparing its results for the semi-major axis with those obtained from different resolved visual binary orbits available in the WDS-ORB6. For the 9 cross-matched orbits in both Gaia and ORB6 catalogs, the algorithmic semi-major axes differ, on average, less than $10\%$ from the ground-based ones. For the rest of the cross-matched binaries, the discrepancies in all of their orbital parameters indicate that they correspond to different components in those systems so that the semi-major axes can not be compared, and more observations are needed to evaluate the algorithm in multiple systems. 

We hope that this work will serve as a complementary tool to exploit the huge amount of astrophysical data provided by Gaia. A more complete version of this draft is expected to be ready for publication soon. 

\section{Acknowledgements}

This work presents results from the European Space Agency (ESA) space mission Gaia. Gaia data are being processed by the Gaia Data Processing and Analysis Consortium (DPAC). Funding for the DPAC is provided by national institutions, in particular the institutions participating in the Gaia Multi-Lateral Agreement (MLA). The Gaia mission website is \url{https://www.cosmos.esa.int/gaia}. The Gaia Archive website is \url{http://archives.esac.esa.int/gaia}.

Another catalogs and databases that have been crucially important for the preparation of this work, are: the Washington Double Star Catalog (WDS) and the Sixth Catalog of Orbits of Visual Binary Stars (ORB6), maintained by the US Naval Observatory (USA) (\url{https://crf.usno.navy.mil/wds}); SIMBAD database, operated by the \textit{Centre de Données astronomiques de Strasbourg} (France) (\url{http://simbad.cds.unistra.fr/simbad/}); and the Astrophysics Data System (NASA/SAO) (\url{https://ui.adsabs.harvard.edu/}).

This paper was supported by the Spanish ``Ministerio de Ciencia e Innovación'' under the Project PID2021-122608NB-I00 (AEI/FEDER, UE).


\end{document}